\documentclass[lettersize,journal]{IEEEtran}
\usepackage{amsmath,amsfonts}
\usepackage{algorithmic}
\usepackage{algorithm}
\usepackage{array}
\usepackage[caption=false,font=footnotesize,labelfont=rm,textfont=rm]{subfig}
\usepackage{textcomp}
\usepackage{stfloats}
\usepackage{url}
\usepackage{verbatim}
\usepackage{graphicx}
\usepackage{cite}
\usepackage{stfloats}
\usepackage{makecell}
\usepackage{multirow}
\usepackage{amsthm,amsmath,amssymb}
\usepackage{mathrsfs}
\usepackage{epsfig}
\usepackage{epstopdf}
\usepackage{diagbox}
\usepackage{soul}
\usepackage{ulem}
\usepackage{color, xcolor}
\usepackage{pifont}
\begin{document}

\title{A Novel Multi-Reference-Point Modeling Framework for Monostatic Background Channel: Toward 3GPP ISAC Standardization}

\author{Yameng Liu, Jianhua Zhang, Yuxiang Zhang, Zhiqiang Yuan, Chuangxin Jiang, \\Junchen Liu, Wei Hong, Yingyang Li, Yan Li, Guangyi Liu\\
\thanks{This research is supported in part by National Natural Science Foundation of China under Grant 62525101, Grant 62201087, in part by the National Key R\&D Program of China under Grant 2023YFB2904803, in part by the Guangdong Major Project of Basic and Applied Basic Research under Grant 2023B0303000001, in part by the Natural Science Foundation of Beijing-Xiaomi Innovation Joint Foundation under Grant L243002, and in part by the Beijing University of Posts and Telecommunications-China Mobile Research Institute Joint Institute.}
\thanks{Yameng Liu, Jianhua Zhang, Yuxiang Zhang are with the State Key Laboratory of Networking and Switching Technology, Beijing University of Posts and Telecommunications, Beijing 100876, China (email: liuym@bupt.edu.cn; jhzhang@bupt.edu.cn; zhangyx@bupt.edu.cn).}
\thanks{Zhiqiang Yuan is with the National Mobile Communications Research Laboratory, Southeast University, Nanjing 210096, China (email: zqyuan@seu.edu.cn)} 
\thanks{Chuangxin Jiang and Junchen Liu are with the algorithm department, ZTE corporation, Xi'an 710065, China (email: jiang.chuangxin1@zte.com.cn, liu.junchen@zte.com.cn)} 
\thanks{Wei Hong and Yingyang Li are with the Xiaomi Industry Standardization and Research Department, Beijing 100085, China (e-mail: hongwei@xiaomi.com, liyingyang@xiaomi.com).}
\thanks{Yan Li is with the Department of Wireless and Device Technology Research, China Mobile Research Institute, Beijing 100053, China (e-mail: liyanwx@chinamobile.com).}
\thanks{Guangyi Liu is with the Future Research Laboratory, China Mobile Research Institution, Beijing 100053, China, (email: liuguangyi@chinamobile.com).}}

\markboth{Journal of \LaTeX\ Class Files,~Vol.~14, No.~8, August~2021}
{Shell \MakeLowercase{\textit{et al.}}: A Sample Article Using IEEEtran.cls for IEEE Journals}


\maketitle

\begin{abstract}
Integrated Sensing and Communication (ISAC) has been identified as a key 6G application by ITU and 3GPP. A realistic, standard-compatible channel model is essential for ISAC system design. To characterize the impact of Sensing Targets (STs), 3GPP defines ISAC channel as a combination of target and background channels, comprising multipath components related to STs and those originating solely from the environment, respectively. Although the background channel does not carry direct ST information, its accurate modeling is critical for evaluating sensing performance, especially in complex environments. Existing communication standards characterize propagation between separated transmitter (Tx) and receiver (Rx). However, modeling background channels in the ISAC monostatic mode, where the Tx and Rx are co-located, remains a pressing challenge. 
In this paper, we firstly conduct ISAC monostatic background channel measurements for an indoor scenario at 28 GHz. Realistic channel parameters are extracted, revealing pronounced single-hop propagation and discrete multipath distribution. Inspired by these properties, a novel stochastic model is proposed to characterizing the ISAC monostatic background channel as the superposition of sub-channels between the monostatic Tx$\&$Rx and multiple communication Rx-like Reference Points (RPs). This model is compatible with standardizations, and a 3GPP-extended implementation framework is introduced. 
Finally, a genetic algorithm–based method is proposed to extract the optimal number and placement of multi-RPs. The optimization approach and modeling framework are validated by comparing measured and simulated channel parameters. Results demonstrate that the proposed model effectively captures monostatic background channel characteristics, addresses a critical gap in ISAC channel modeling, and supports 6G standardization.

\end{abstract}


\section{Introduction}\label{section1}
\IEEEPARstart{I}{n} Dec. 2023, Integrated Sensing and Communication (ISAC) was identified as one of the key applications of sixth generation (6G) by International Telecommunication Union (ITU) \cite{itu6G,liu2024cooperative}. Concurrently, Third Generation Partnership Project (3GPP) Radio Access Network (RAN) officially initiated a study item on ISAC channel modeling \cite{3gppRan102}. ISAC research has attracted extensive attention and exploration from both industry and academia \cite{zhang2024latest}. Compared with conventional systems employing separate devices, ISAC enables communication Base Stations (BSs) or User Terminals (UTs) to simultaneously sense the surrounding environment \cite{zhang2018multibeam,pucci2022system}. This integration shows great potential for improving spectrum efficiency and reducing system costs \cite{liu2023shared,noh2022communication}. 

A realistic, accurate, and standard-compatible channel model is essential for the deployment and evaluation of 6G ISAC technologies. The mainstream Geometry-Based Stochastic Model (GBSM) has been widely adopted in standardized communication channel modeling by ITU and 3GPP \cite{3gpp38901,itu2412}. It is two-dimensions (2D) in 4G systems and expands to 3D in 5G systems by introducing the zenith angles \cite{zhang20173}. GBSM
adopts a clustered structure, grouping multipaths into stochastic clusters. In ISAC systems, effective evaluation of sensing tasks, such as target localization and tracking, requires channel models to incorporate deterministic representations of Sensing Targets (STs) \cite{zhang2025unifiedRCS,yuan2025experimental}. However, the overall propagation characteristics in ISAC channels still follow statistical patterns, making GBSM a suitable foundation for model extension \cite{liu2022survey,gong2024new,kumari2017ieee}. This consensus has already been reached in initial 3GPP discussions \cite{3gppRan102}. 
To facilitate accurate sensing evaluations, the ISAC channel framework has further been defined as a combination of target and background channels, comprising multipath components related to the STs and those originating solely from the environment, respectively. This definition was formally adopted at the 3GPP RAN Working Group 1 (RAN1) meeting in Feb. 2024 \cite{3gppRan116FL}.

\begin{figure}[h]
\centering
\includegraphics[width=3.4in]{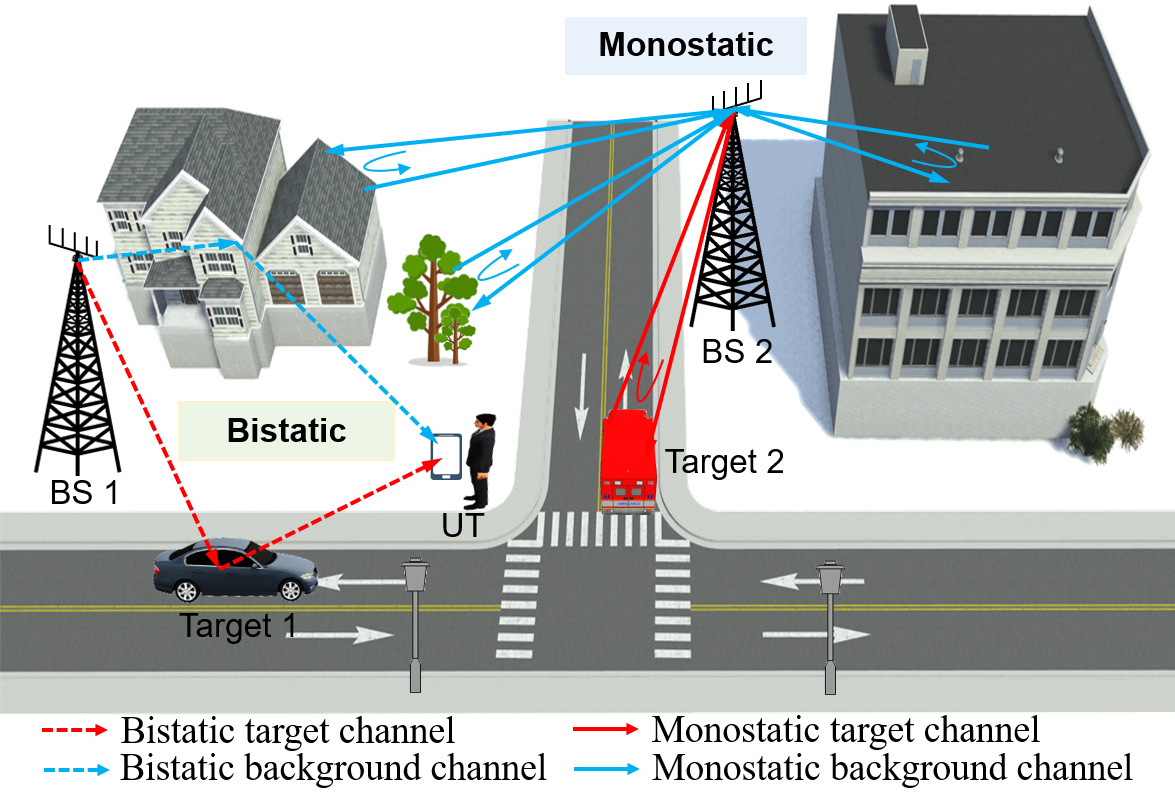}
\caption{The illustration of ISAC channel. Red and blue dashed lines denote the bistatic target and background channels, respectively. Red and blue solid lines denote the monostatic target and background channels, respectively.}
\label{fig_question}
\end{figure}

Fig. \ref{fig_question} illustrates a typical example of ISAC channel. The target channel (denoted by the red lines) involves transmitter (Tx)-ST-receiver (Rx) propagation, where the ST serves as an anchor with a known position and velocity, comparable to the Tx or Rx. Its characteristics have been extensively studied in existing literature, e.g., \cite{zhang2025research, abadpour2022angular,zhang2025unifiedRIS}. However, the background channel multipaths (denoted by the blue lines in Fig. \ref{fig_question}) actually constitute a larger fraction of the ISAC channel, which has been practically validated in \cite{zhang2025research}. The work in \cite{liu2025coupling} highlights the importance of the background channel and analyzes the coupling effects between the background multipath components and the ST. \cite{wang2020small} investigates small target tracking in satellite videos by introducing background information into training. In theoretical analysis and design, simplified ideal assumptions about the background channel have also been initially applied. For example, in \cite{xiong2023fundamental,xiong2024torch}, the ST-unrelated component is considered as additive white Gaussian noise, with ISAC system performance derived accordingly. Similar channel assumptions are employed in \cite{yu2022location} and \cite{chen2021code}, where beamforming and system design are explored, respectively. Although the background channel does not carry direct ST information, its accurate modeling is crucial for evaluating sensing performance, especially in complex environments \cite{zhang2021overview,liu2025coupling}. 

ISAC background signals exhibit two propagation modes \cite{liu2022survey}: bistatic and monostatic, as shown in Fig. \ref{fig_question}. The bistatic background channel, represented by the blue dashed lines, involves a spatial separation between the Tx, i.e., BS 1, and Rx, i.e., UT, similar to communication configurations. The conventional GBSM generates propagation channels based on the relative positions of the transceivers as well as statistical parameter distributions \cite{3gpp38901}, making it readily applicable to bistatic background sensing \cite{luo2024channel,yang2024integrated}. 
In contrast, the blue solid lines in Fig. \ref{fig_question} depict the monostatic background channel, where the co-located Tx and Rx (Tx$\&$Rx), i.e., BS 2, acquire channel information by receiving its own transmitted signals affected by environmental scatterers. The absence of independent anchors (e.g., Rx or ST) relative to the Tx renders existing GBSMs are not applicable. While radar research \cite{lampropoulos1999high,addabbo2021learning} has modeled monostatic clutter using statistical distributions, it fails to achieve compatibility with communication standards, making it challenging to perform unified performance evaluations for ISAC systems. 

To capture realistic characteristics of the monostatic background channel, state-of-the-art studies \cite{zhang2025research, chen2024empirical, ali2020leveraging} have conducted several measurements, demonstrating that propagation parameters are influenced by the scenario conditions, such as the density of environmental scatterers around the Tx$\&$Rx. In \cite{barneto2022millimeter}, a geometric model is proposed based on an assumed small distance between the monostatic Tx and Rx. However, the high complexity of this approach limits its applicability for standardization. To enhance compatibility with existing standards, \cite{jiang2024novel} proposes an interesting virtual communication Rx approach, where the channel from the Tx$\&$Rx to the single virtual Rx, which can be referred to as a Reference Point (RP), is modeled by reusing the 3GPP Technical Report (TR) 38.901 \cite{3gpp38901}.
In this work, monostatic Ray-tracing (RT) simulations are applied, and the strongest path is identified as the RP position. However, measurement results in \cite{liu2023shared} indicate that monostatic background channels exhibit a more discrete multipath distribution than that of communication channels, with significantly larger Delay Spread (DS) and Angular Spread (AS), parameters that traditional GBSMs with single Rx (i.e., single RP) fail to capture. As a result, accurately and practically developing an extended GBSM framework for ISAC monostatic background channels remains an open challenge, highlighting the urgent need for further methodological advancements and more validation.

To bridge the above gaps, we have conducted an indoor ISAC measurement campaign to obtain the realistic channel propagation characteristics. A novel multi-RP modeling framework is proposed, analyzed theoretically, and validated based on the measured parameters. Our major contributions and novelties are summarized as:

\begin{itemize}
\item{An ISAC monostatic background channel measurement campaign is conducted at 28 GHz in a typical indoor scenario, using two co-located horn antennas with 5° rotational steps to capture realistic channel characteristics. The channel paths are observed to exhibit pronounced single-hop propagation and a discrete distribution, as analyzed from the measured Power-Angular-Delay Profile (PADP).} Channel propagation parameters, including Path Loss (PL), DS, and AS, are extracted to provide a data foundation for modeling development. 

\item{A realistic, standard-compatible multi-RP modeling framework for ISAC monostatic background channels is proposed. Specifically, the multi-RP modeling principle is analyzed based on the observed channel characteristics, motivating the representation of the monostatic background channel as the superposition of sub-channels between Tx$\&$Rx and multi-RPs. A 3GPP-extended channel implementation framework is proposed, ensuring standardization compatibility and enabling low-complexity realization of ISAC monostatic background channels.}

\item{The parameterization and validation of the proposed channel model are conducted based on measured data. A Genetic Algorithm-based Multi-RP Extraction (GA-MRPE) method is proposed to determine the optimal multi-RPs. This algorithm identifies 3 multi-RPs with approximately uniformly distributed horizontal angles. Results validate that the proposed model accurately matches the measured ISAC channel, with parameter errors reduced by over 20\% and 50\% in the delay and angular domains, respectively, compared to reusing communication channel parameters.}
\end{itemize}

The remainder of this paper is outlined as follows. In Section \ref{section2}, detailed descriptions of the measurement facilities and environment are presented, and the channel parameters are extracted. In Section \ref{section3}, a multi-RP channel model and a 3GPP-extended implementation framework are proposed for modeling monostatic background channels. Then, a GA-MRPE method for model parameterization is proposed, and the validation of the proposed model is accomplished in Section \ref{section4}. Finally, Section \ref{section5} concludes the work.

\section{Monostatic Background Channel Measurements}\label{section2}

\subsection{Measurement Description}

\begin{figure}[h]
\centering
\subfloat[]{\includegraphics[width=3.4in]{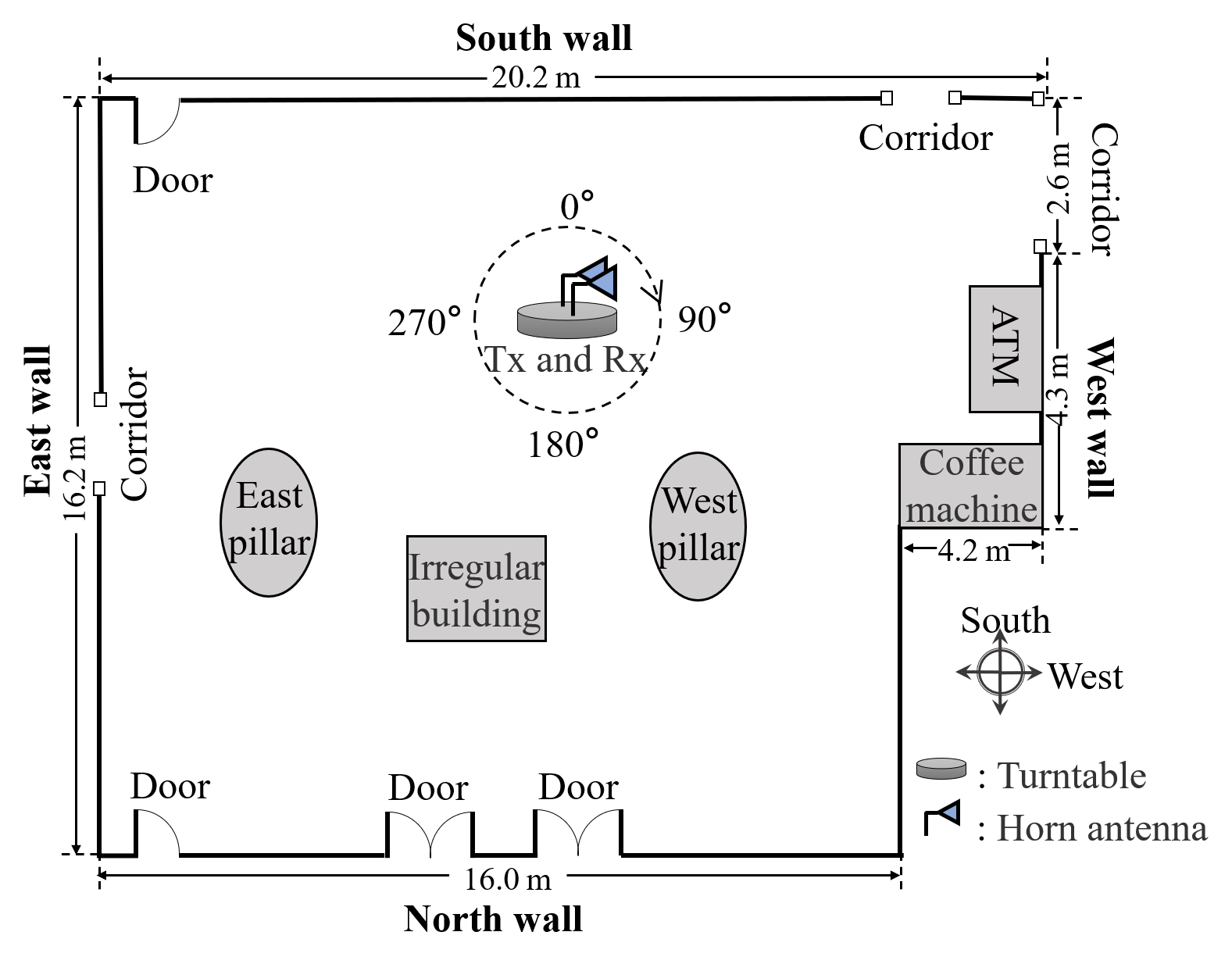}\label{fig_sce1}}\\
\subfloat[]{\includegraphics[width=2.9in]{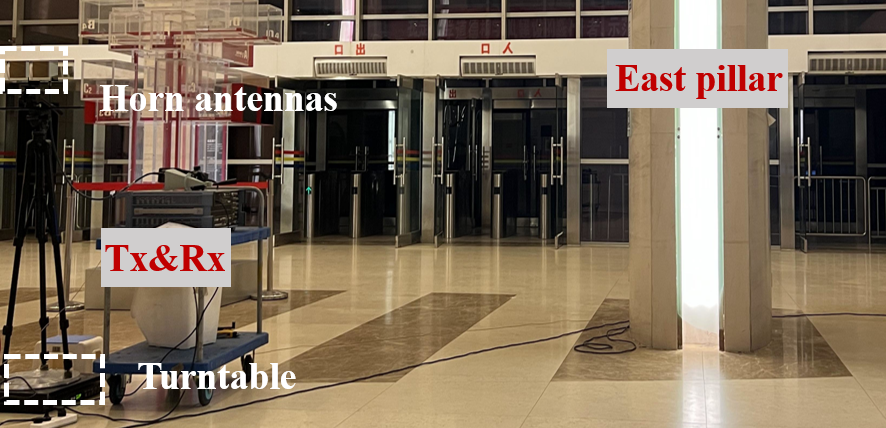}\label{fig_sce2}}\\
\caption{The illustration of (a) the measurement hall layout and (b) a photograph taken at the horizontal angle of 255$^\circ$ in the measurement scenario.}
\label{fig_sce}
\end{figure}

The measurements for ISAC monostatic background channels without STs are conducted at 28 GHz in a typical indoor hall, with the dimension of 20.2$\times$16.2 $\text{m}^2$ at Beijing University of Posts and Telecommunications. The scenario layout and the actual measurement surroundings are shown in Fig. \ref{fig_sce}.

In the measurements, the Tx and Rx sides are equipped with a horn antenna, respectively. These two antennas are fixed on a bracket with a horizontal interval of 10 cm, pointing in the same direction to detect the monostatic propagation path of sensing signals. Absorptive materials are deployed between the Tx and Rx antennas to physically block the leakage of the Line-of-Sight (LoS) path. (In monostatic configurations, the LoS path carries no environmental information and primarily contributes to self-interference in the ISAC system.) The antenna bracket is mounted on a turntable located at the center of the hall, as indicated by the circular dashed area in Fig. \ref{fig_sce1}. The height of monostatic Tx$\&$Rx antennas are set at a height of 1.47 m. To obtain omni-directional channel information, the antennas are rotated horizontally in 5$^\circ$ increments, covering a full 360$^\circ$ clockwise sweep starting from the south, resulting in 72 angular measurements. (This directional scanning method and rotation step size have been validated \cite{fan2020angular,hu2017effects} to capture channel characteristics within an acceptable error margin, while reducing costs and measurement overhead, contributing to its widespread use.)
The vertical angle is fixed at 90$^\circ$. It is important to note that the use of closely spaced dual antennas in this setup emulates the co-located transceivers in the ISAC monostatic configuration, without compromising the general applicability of the channel analysis.

\begin{figure}[h]
\centering
\includegraphics[width=2.8in]{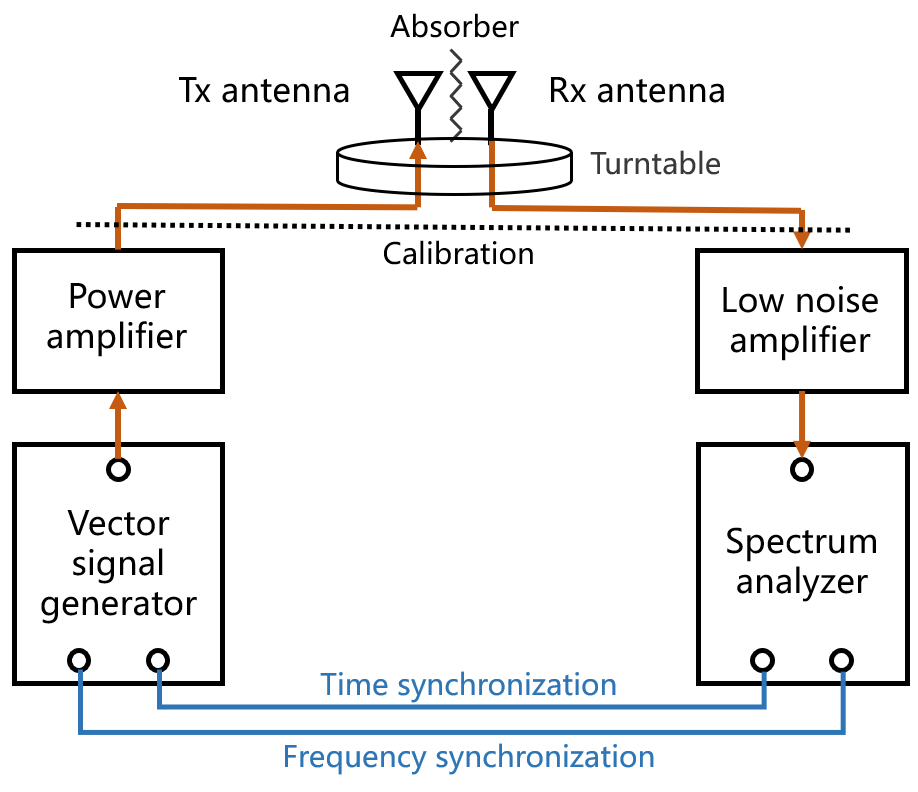}
\caption{A time-domain correlation based channel sounder at mmWave bands, which mainly consists of a vector signal generator, a power amplifier, a low-noise amplifier, a spectrum analyzer, and two horn antennas. Calibration between the Tx and Rx is performed before measurements to de-embed the system response.}
\label{fig_config}
\end{figure}

A wideband channel sounder operating at millimeter wave (mmWave) bands is employed to extract the ISAC channel characteristics, as illustrated in Fig. \ref{fig_config}. At the Tx side, a vector signal generator (R$\&$S SMW 200A) is used to generate a Pseudo Noise (PN) sequence with a code rate of 500 Msym/s and a length of 511. This baseband PN sequence is then modulated as a mmWave probing signal at 28 GHz using Binary Phase Shift Keying (BPSK) modulation. The zero-to-zero bandwidth is 1 GHz. To improve the Signal-to-Noise Ratio (SNR), this signal is amplified by a power amplifier with 25 dBm saturation power and 35 dB gain, and transmitted via a high-gain horn antenna. At the Rx side, the received signal is captured by a horn antenna, amplified by a Low-Noise Amplifier (LNA) providing 20 dB gain, and then down-converted using a spectrum analyzer (R$\&$S FSW 43). A total of 1022 I/Q samples are collected at a sampling rate of 1 GHz, corresponding to a delay resolution of 1 ns. The Channel Impulse Responses (CIRs) are then obtained through offline processing on a laptop.

In the channel measurements, multiple channel samplings are realized, which are expressed as sampling cycles (snapshots). Each snapshot has a maximum detectable delay of 1022 ns, corresponding to a maximum sounding distance of 306.6 m. Moreover, to obtain synchronized delay and frequency, we directly connect the spectrum analyzer and the signal generator through cables during the measurements. Details of the channel sounder configuration are provided in Table \ref{table_1}.

\begin{table}[h]
\caption{Channel Sounder Configuration Parameters\label{table_1}}
\centering
\begin{tabular}{cc}
\hline
Parameter & Values\\
\hline
Central frequency & 28 GHz\\
Symbol rate & 500 Msym/s\\
Bandwidth & 1 GHz\\
PN sequence & 511\\
Sampling rate & 1 GHz\\
Tx / Rx antenna type & Horn / Horn\\
Antenna azimuth HPBW  & 10$^\circ$ \\
Antenna vertical HPBW  & 8$^\circ$ \\
Antenna gain & 25 dBi\\
Antenna height & 1.47 m\\
Power amplifier gain & 35 dB\\
Low-noise amplifier gain & 20 dB\\
\hline
\end{tabular}
\end{table}

\subsection{Channel Characteristic Analysis} 
\begin{figure}[h]
\centering
\includegraphics[width=2.9in]{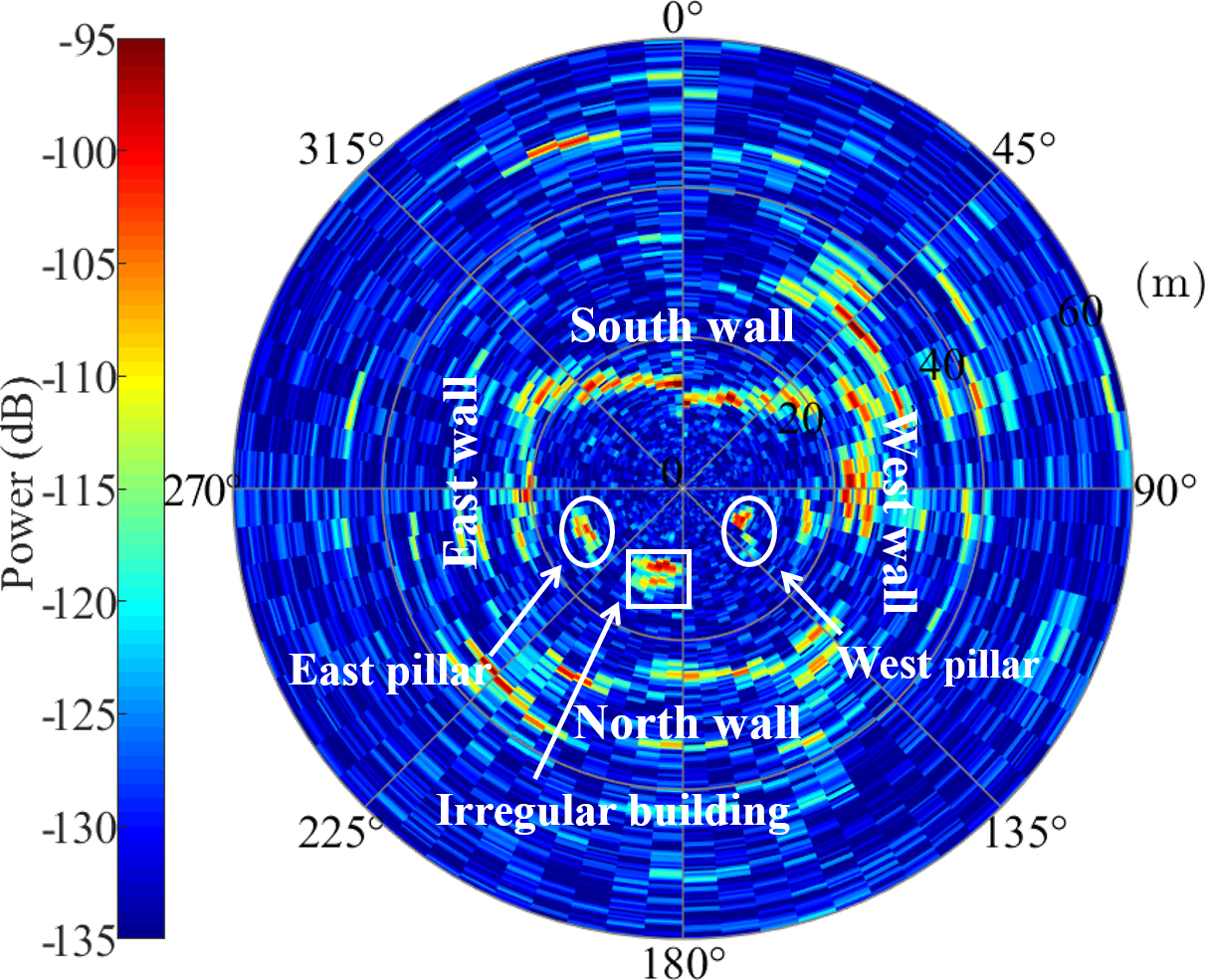}
\caption{Measured PADP$(\theta,\tau)$ for ISAC monostatic background channel. The PADP center represents the location of the monostatic Tx$\&$Rx antennas.}
\label{fig_padp}
\end{figure}

To remove the system response from the equipment, antennas, and cables, we perform a back-to-back calibration on the field measurement data \cite{jiang2019comparative}. CIRs of monostatic background channel for all rotation angles is measured as $h(\theta,\tau)$, where $\tau$ and $\theta$ denote the propagation delay and rotation angle, respectively. The Power-Angular-Delay Proﬁle (PADP) of ISAC monostatic background channel can be expressed as
\begin{equation}
\label{eqn_padp}
\text{PADP}(\theta,\tau)=|h(\theta,\tau)|^2,
\end{equation}
which is demonstrated in Fig. \ref{fig_padp}. In the polar plot, the center of the circle (coordinate of $(0,0)$) represents the location of the Tx$\&$Rx antennas, as shown in Fig. \ref{fig_sce1}. The angle of the circle represents the Azimuth angle of Departure (AoD) of paths, with zero-degree pointing in the south direction. The radius indicates the absolute propagation distance from the Tx antenna to the Rx antenna, and the depth of color represents the magnitude of the received power (dB). (Note that the received power values are not normalized here.) By matching the high-power paths with the actual environment, it can be calculated that around 90\% of the path power comes from the single-hop reflection, within the 30 dB dynamic range of the measured PADP.

Considering the round-trip characteristics of the monostatic channel paths, the measured single-hop propagation distance is twice the actual distance between the scatterer and the Tx$\&$Rx antennas. As a result, the absolute delay and AoD of the monostatic background paths can be used to determine the location of environmental scatterers, enabling localization. As demonstrated in Fig. \ref{fig_padp}, the scenario layout, including the south, east, north, and west walls of the hall, the east and west pillars, and the irregularly shaped building, is clearly presented and corresponds to Fig. \ref{fig_sce1}. For multi-hop paths, the angle between the Tx antenna and the first-hop scatterer is always intuitively reflected by the AoD of the paths in the PADP. However, the specific propagation route cannot be determined solely from the available delay and angle parameters. Moreover, these multi-hop paths distribute more scattered at farther locations. For example, in areas like corridors and doorways (e.g., around 45$^\circ$ and 40 m in Fig. \ref{fig_padp}), paths are randomly distributed due to multiple reflections, diffraction, etc., caused by irregular scatterers.

\begin{figure}[h]
\centering
\subfloat[]{\includegraphics[width=2.5in]{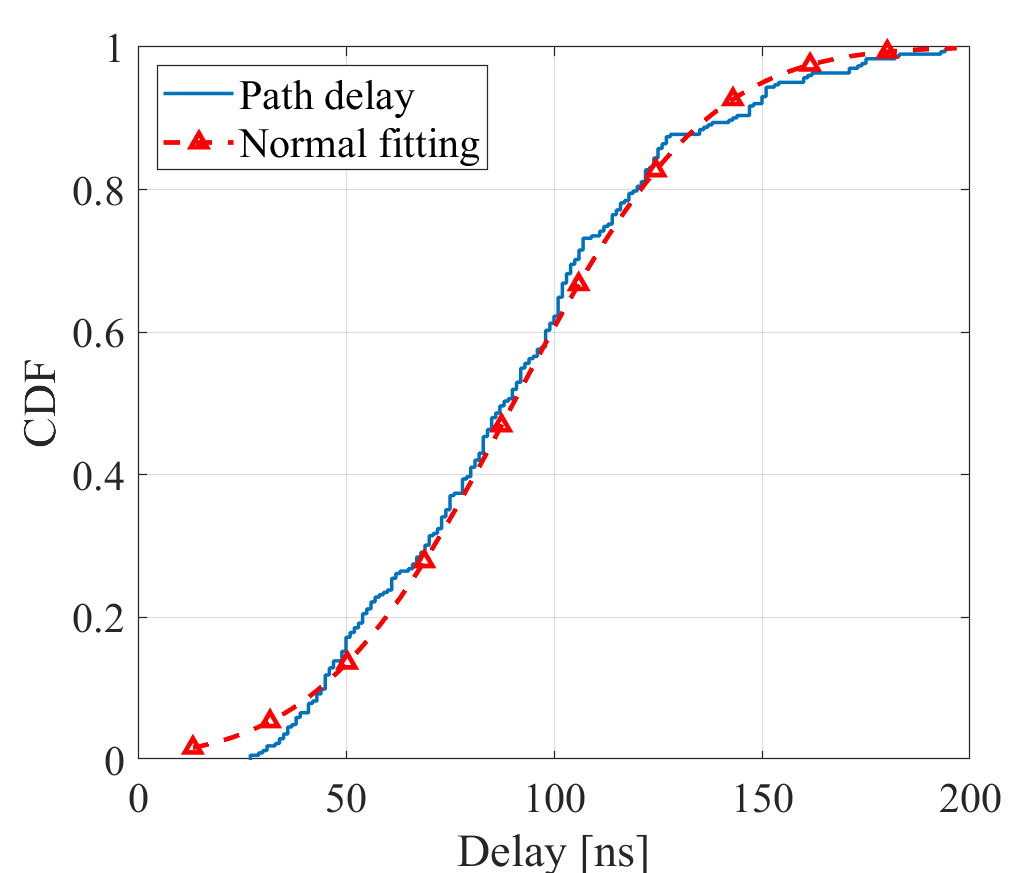}
\label{fig_meaT22}}\\
\vspace{-4mm}
\subfloat[]{\includegraphics[width=2.5in]{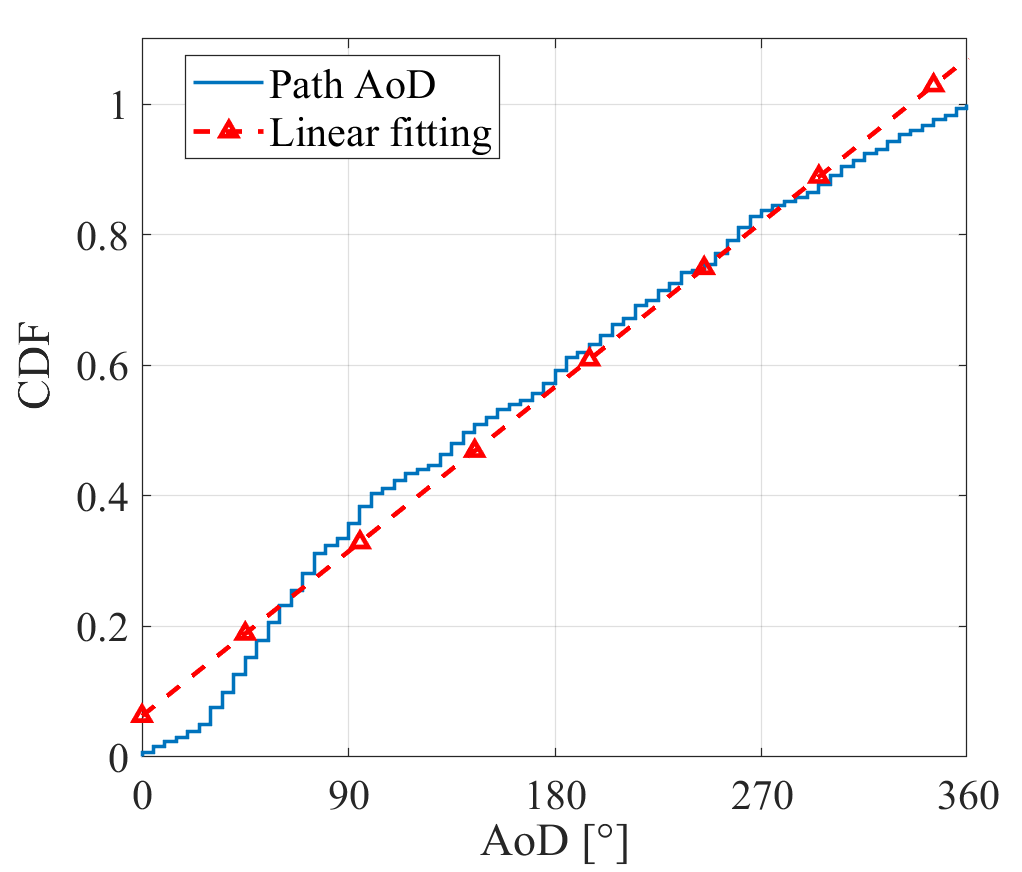}
\label{fig_meaA2}}\\
\caption{Measured CDFs of propagation parameters in ISAC monostatic background channel, including (a) path delay [ns], (b) path AoD [$^\circ$].}
\label{fig_meapara}
\end{figure}

Here, the monostatic background channel incorporates all environmental scatterers, meaning that waves propagating to the scatterers always return a portion of the energy that can be received. This is due to the high detection dynamic range provided by the channel sounder gain and the strong diffuse reflection mechanisms facilitated by the rough surfaces. As a result, the distribution of the effective paths is nearly 360$^\circ$ omnidirectional. The Cumulative Distribution Functions (CDFs) of effective path delays and angles are drawn in Fig. \ref{fig_meapara}. In this scenario and frequency, the measured delays follow a normal distribution with a mean of 90.20 ns and a standard deviation of 36.39 ns, while the measured AoDs follow a uniform distribution ranging from 0$^\circ$ to 360$^\circ$.

Based on these measurement results, we further calculate the channel stochastic parameters, including the channel PL, the root mean square DS (rms DS), and the rms AS of AoDs.
Specifically, the power of each valid multipath component is extracted, and the PL of the ISAC monostatic background channel is then approximated by summing the linear power values ($p_m$) of these components, which is typically converted into the dB domain. The calculation is expressed as
\begin{equation}
\label{eqn_meapl}
PL_0[{\rm{dB}}]=10\cdot\log_{10}\left(\sum_{m=1}^{M}{p_m}\right),
\end{equation}
where $M$ is the total number of the effective paths, with $M=302$ here. The estimated PL of the measured monostatic background channel in this paper is -80.8125 dB. Note that, this is merely a numerical issue in different scenarios and does not affect our subsequent exploration of the modeling methodology for ISAC monostatic background channel.
Moreover, the rms DS ($\sigma_{\tau}$) of all the effective monostatic channel paths is calculated as
\begin{gather}
\label{eqn_spru}
\mu_{\tau}=\left(\sum\limits_{m=1}^{M}\tau_m\cdot p_m\right)/\sum\limits_{m=1}^{M}p_m,\\
\label{eqn_sprr}
\sigma_{\tau}=\sqrt{\sum\limits_{m=1}^{M}(\tau_m-\mu_{\tau})^2\cdot p_m/\sum\limits_{m=1}^{M}p_m},
\end{gather}
where $\mu_{\tau}$ indicates the mean delay of the effective paths. $\tau_m$ and $p_m$ are the delay and power of the path $m$, respectively. Replacing $\tau$ with $\theta$ in the above formulas, the azimuth AS can be derived in the same manner. 
As for the measured monostatic background channel in this paper, the calculated DS ($DS_0$) is 32.92 ns, and the AS ($AS_0$) is 89.98$^\circ$, indicating a relatively discrete multipath structure. For comparison, the corresponding mean values in the indoor NLoS scenario at 28 GHz in the communication GBSM are 26.15 ns and 41.69$^\circ$. These realistic channel propagation parameters will provide valuable data support for modeling development.

\section{Multi-Reference-Point based Monostatic Background Channel Modeling for ISAC systems}\label{section3}

\begin{figure*}[t]
\centering
\includegraphics[width=7.2in]{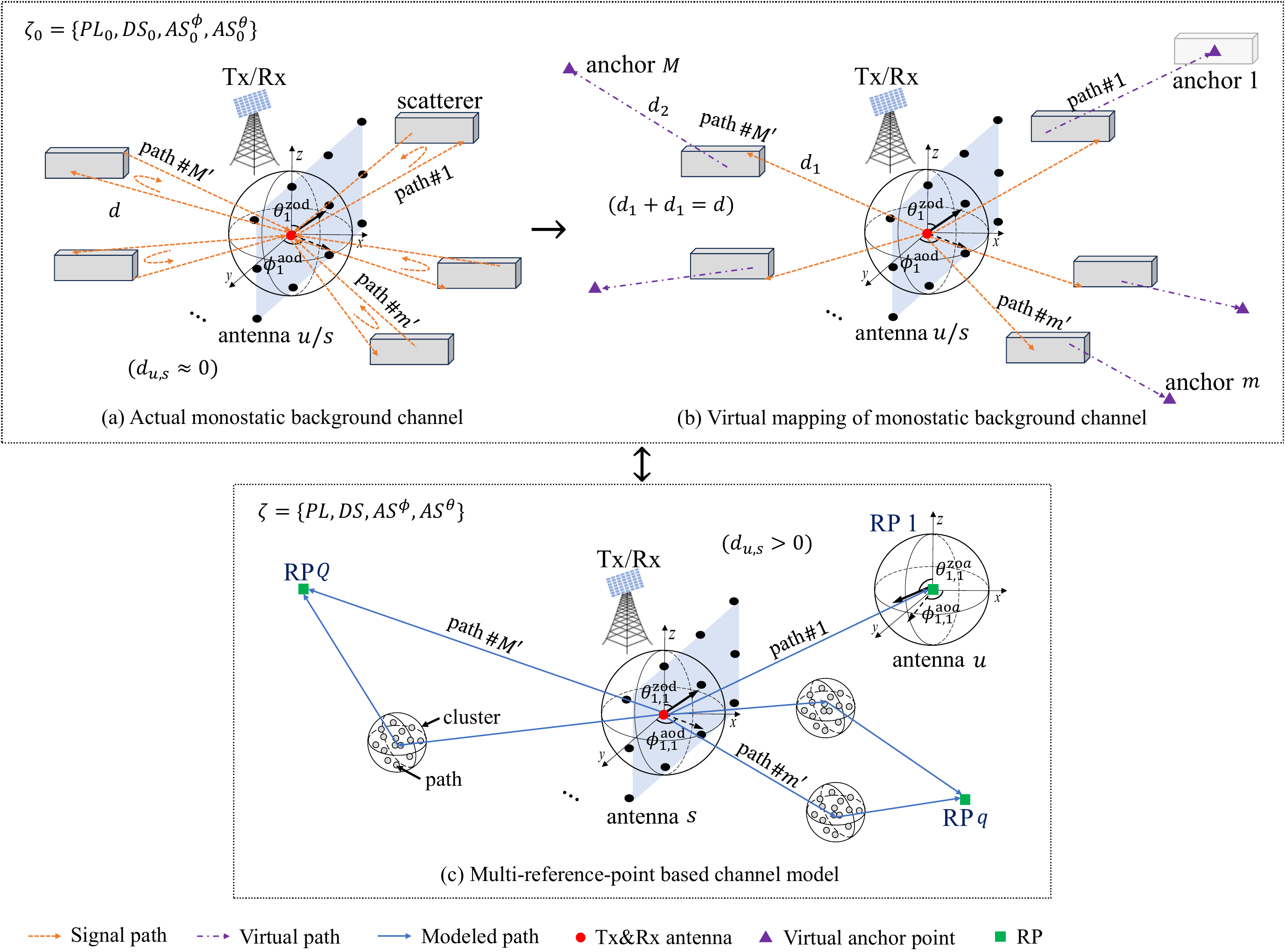}
\caption{Illustration of the proposed model based on multi-RPs for the ISAC monostatic background channel. The orange and purple dashed lines denote the realistic and virtual monostatic paths, respectively. The blue solid lines denote the modeled paths. The red dots represent the monostatic Tx$\&$Rx antenna, the purple triangles denote the virtual anchor points, and the green squares indicate the modeled multi-RPs.}
\label{fig_modelmr}
\end{figure*}

An important objective in developing an accurate stochastic channel model and its simulation implementation is to ensure that the modeled channel statistical parameters, such as ${\boldsymbol{\zeta}}=\{PL,DS,AS^\phi,AS^\theta\}$ (i.e., PL ($PL$), DS ($DS$), azimuth AS ($AS^\phi$), and zenith AS ($AS^\theta$)) closely align with empirical measurement results, such as ${\boldsymbol{\zeta}}_0=\{PL_0,DS_0,AS_0^\phi,AS_0^\theta\}$. This alignment serves as a key criterion for evaluating model accuracy. In this section, we  extend the 3D GBSM \cite{3gpp38901} and propose a multi-Reference-Point (multi-RP) based modeling framework to characterize the monostatic background channel for ISAC systems. This framework trends the modeled parameter set (${\boldsymbol{\zeta}}$) toward the actual measurement values (${\boldsymbol{\zeta}}_0$).

\subsection{Multi-Reference-Point based Channel Modeling Principle}

This paper considers a wideband ISAC monostatic background channel with $S$ Tx antenna array elements and $U$ Rx antenna array elements. The $s^{th}$ and $u^{th}$ elements may either be the same antenna or closely located antennas.
In the practical monostatic background channel, the signal is transmitted from the Tx$\&$Rx, undergoes reflection, scattering, or diffraction by surrounding scatterers, and is then received back at the Tx$\&$Rx as an echo. This propagation is illustrated by the orange dashed lines in Fig. \ref{fig_modelmr}(a), with the gray squares representing the propagation scatterers distributed throughout the environment. Note that a single scatterer may contribute to multiple propagation paths, with the paths depicted in the Fig. \ref{fig_modelmr} serving as illustrations only. The actual channel propagation parameter set is denoted as $\boldsymbol{\zeta}_0$. As discussed in Section \ref{section1}, the standardized GBSM characterizes propagation characteristics between separated transceivers. For instance, the calculation of channel PL and absolute cluster / path delays depends on the Tx-Rx distance ($d_{u,s}$), while the cluster / path azimuth and zenith angles are determined by their relative positions. Therefore, GBSM is inapplicable for monostatic background channels, as the Tx-Rx distance is approximately zero ($d_ {u,s} \approx 0$).

Nevertheless, monostatic sensing can still be regarded as a special communication case, following electromagnetic propagation and exhibiting consistent propagation parameters. Therefore, the modeling of the monostatic background channel can also be approached from a communication perspective to develop a unified ISAC model. Inspired by the observations in Section \ref{section2}-B, which indicate that most effective multipaths in the ISAC monostatic background channel are single-hop, the actual positions of the monostatic scatterers can be extended up to twice the distances to form virtual anchor points. These extended paths and anchor points are indicated by the purple dashed lines and triangles in Fig. \ref{fig_modelmr}(b). As a result, the round-trip distance from the monostatic Tx$\&$Rx to the scatterer ($d$ shown in Fig. \ref{fig_modelmr}(a)) is equivalent to the one-way distance to the virtual anchor point ($d_1+d_2$ shown in Fig. \ref{fig_modelmr}(b)). (Note that considering multi-hop monostatic path, a virtual anchor point can still be found, aligned with the path departure angle and propagation delay, but it cannot be directly determined from the scatterer position.) The propagation parameters of the monostatic background path $\{1,...,m',...,M'\}$ can be characterized based on the corresponding virtual anchor point.

However, as observed in the measurements in Section \ref{section2}-B, the multipath characteristics of the monostatic background channel exhibit significant discreteness, leading to a large number of effective paths and virtual anchor points. This is especially true in ISAC systems, where high resolution is required to support sensing performance. Modeling all the links from the Tx$\&$Rx to each anchor point using GBSM would introduce substantial complexity, undermining the advantages of stochastic models. Therefore, we consider introducing simplifications by treating the path of an anchor point as a component of the channel from the Tx$\&$Rx to an RP, with the paths generated by adjacent anchor points collectively forming an RP-related channel, as shown in Fig. \ref{fig_modelmr}(c). (Since the standardized GBSM inherently consists of multiple clusters and paths, this assumption is feasible.) The number of RPs, denoted by $Q$, will be much smaller than the number of anchor points, $M'$.

Therefore, this paper proposes a multi-RP based stochastic channel model, which characterizes the ISAC monostatic background channel as the superposition of sub-channels between the Tx$\&$Rx and the multi-RPs. The antenna configuration of each RP matches that of the monostatic Rx $u$ ($d_ {u,s} \textgreater 0$). The propagation parameters of each sub-channel follow the standardized GBSM. As illustrated in Fig. \ref{fig_modelmr}(c), multi-RPs are labeled as \{$1,..., q,...,Q$\}, with the blue lines representing the propagation multipaths. The circles with several dots indicate a scattering region (cluster) responsible for a group of paths with similar properties. Moreover, the modeled channel propagation parameter set is denoted as $\boldsymbol{\zeta}$.

\subsection{3GPP-Extended Channel Modeling Framework}

\begin{figure}[h]
\centering
\includegraphics[width=3.45in]{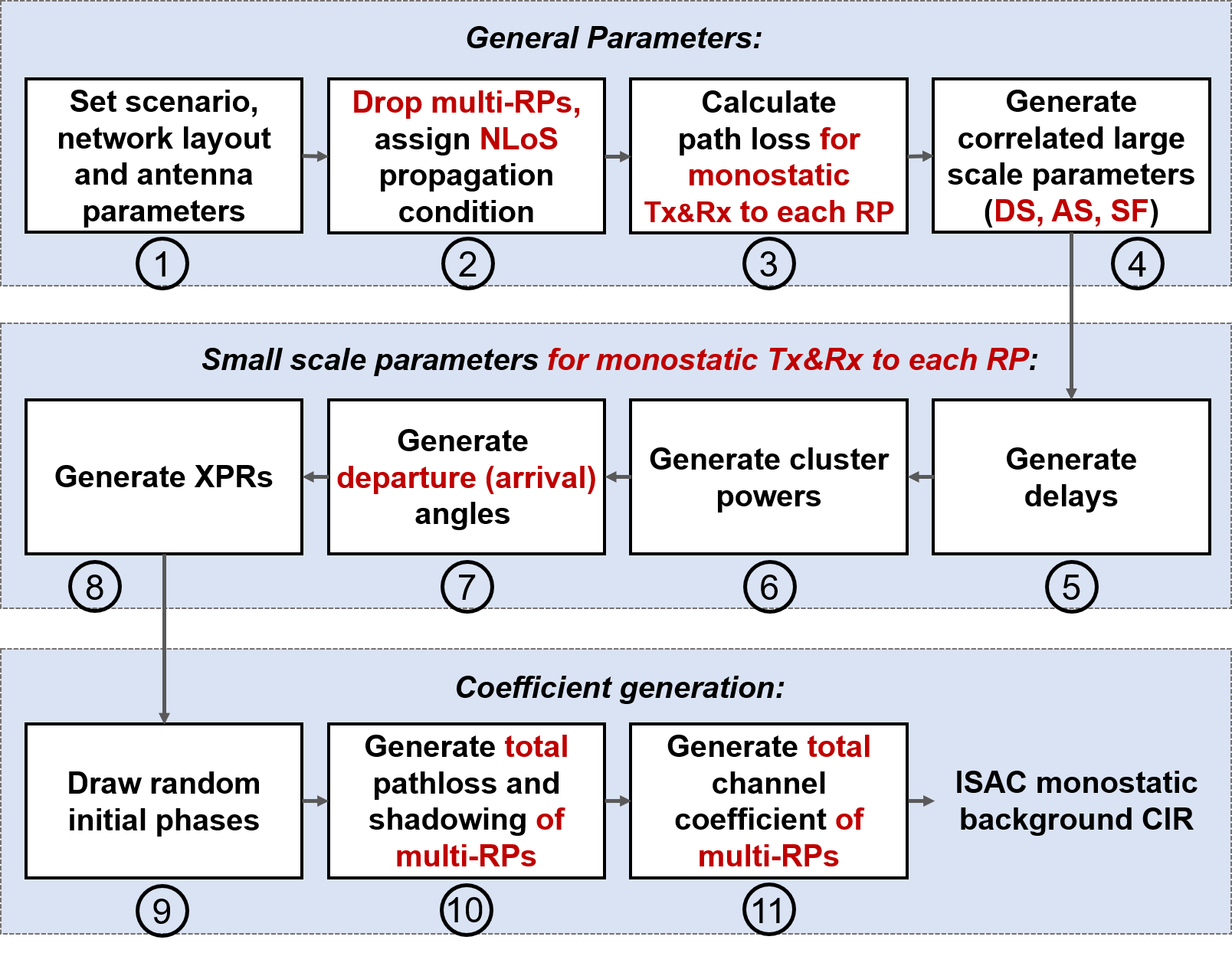}
\caption{Implementation framework of proposed ISAC monostatic background channel model based on multi-RPs. The overall structure follows the 3GPP model \cite{3gpp38901}, with certain steps highlighted in red being modified and adjusted.}
\label{fig_flow}
\end{figure}

Based on the principles of multi-RP channel modeling analyzed above, this paper further proposes a 3GPP-extended modeling framework for ISAC monostatic background channel, as shown in Fig. \ref{fig_flow}.  The overall framework is built upon the 3GPP standardized modeling flow \cite{3gpp38901,zhao2025BUPTCMCC}, with parameter generation steps that deviate from standard specifications are highlighted in red in the diagram.

Specifically, after setting the scenario and antenna parameters in \textit{step 1}, a key issue in the proposed channel model is determining the number and positions of multi-RPs (i.e., as shown in \textit{step 2}). Note that the multi-RPs cannot be directly mapped from the actual environmental scatterers. The model parameterization of multi-RPs will be discussed in subsequent Section \ref{section4}-A.
In the model implementation, we define the delay, AoD, and Zenith angle of Departure (ZoD) of the LoS path in each sub-channel to represent the coordinates of the corresponding multi-RPs. The corresponding expressions are defined as follows:
\begin{gather}
\label{eqn_mrpdist}
{\boldsymbol{\tau}}_{mrp}=\{{\tau_{\rm{los}}^1,...,\tau_{\rm{los}}^q,...,\tau_{\rm{los}}^Q}\},\\
\label{eqn_mrpdisa}
{\boldsymbol{\phi}}_{mrp}^{\rm{aod}}=\{{\phi_{\rm{los}}^{{\rm{aod}},1},...,\phi_{\rm{los}}^{{\rm{aod}},q},...,\phi_{\rm{los}}^{{\rm{aod}},Q}}\},\\
\label{eqn_mrpdisz}
{\boldsymbol{\theta}}_{mrp}^{\rm{zod}}=\{{\theta_{\rm{los}}^{{\rm{zod},1}},...,\theta_{\rm{los}}^{{\rm{zod},q}},...,\theta_{\rm{los}}^{{\rm{zod}},Q}}\}.
\end{gather}
where $\tau_{\rm{los}}^q$, $\phi_{\mathrm{los}}^{\mathrm{aod},q}$ and $\theta_{\mathrm{los}}^{\mathrm{zod},q}$ denote the delay, AoD and ZoD of the LoS path of $q^{th}$ RP.
These parameters, along with the multi-RP number $Q$, should satisfy their corresponding numerical limitations. Moreover, the 3D distances between monostatic Tx$\&$Rx and the multi-RPs are denoted as ${\boldsymbol{d}}_{mrp}^{\rm{3D}}={\boldsymbol{\tau}}_{mrp}\cdot c=\{{d_{rp}^{{\rm{3D}},1},...,d_{{rp}}^{{\rm{3D}},q},...,d_{{rp}}^{{\rm{3D}},Q}}\}$, where $c$ represents the propagation speed of electromagnetic waves. Here, $d_{rp}^{3 \mathrm{D},q}$ is the 3D distance between $q^{th}$ RP and monostatic Tx$\&$Rx. The 3D coordinate of the $q^{th}$ RP ($[{{x}}_{rp}^q,{{y}}_{rp}^q,{{z}}_{rp}^q]$) in the Global Coordinate System (GCS) can be expressed using these distances and angles as
\begin{equation}
\label{eqn_vrcor}
\left(\!\begin{array}{l}
x_{rp}^q \\
y_{rp}^q \\
z_{rp}^q
\end{array}\!\right)=\left(\!\!\begin{array}{l}
x_{tx} \\
y_{tx} \\
z_{tx}
\end{array}\!\!\right)+
\left(\!\begin{array}{l}
d_{rp}^{3 \mathrm{D},q}\sin \theta_{\rm{los}}^{\mathrm{zod},q} \cos \phi_{\rm{los}}^{\mathrm{aod},q} \\
d_{rp}^{3 \mathrm{D},q}\sin \theta_{\rm{los}}^{\mathrm{zod},q} \sin \phi_{\rm{los}}^{\mathrm{aod},q} \\
d_{rp}^{3 \mathrm{D},q}\cos \theta_{\rm{los}}^{\mathrm{zod},q}
\end{array}\!\right),
\end{equation}
where $[x_{tx},y_{tx},z_{tx}]$ is the 3D coordinate of monostatic Tx$\&$Rx.  
The coordinates of all multi-RPs are expressed as $[{\mathbf{x}}_{mrp},{\mathbf{y}}_{mrp},{\mathbf{z}}_{mrp}]_{Q\times 3}$. Moreover, the channel condition is preferably set to Non-LoS (NLoS) to avoid the complex discussion of the K-factor (i.e., the power ratio of the LoS path to the the total power of the NLoS paths) configuration in LoS scenarios.

Among the channel statistical parameters (${\boldsymbol{\zeta}}$), PL is crucial for link budget in engineering applications. 
Given the 3D distance $d_{rp}^{3 \mathrm{D},q}$ from monostatic Tx$\&$Rx to $q^{th}$ RP, the corresponding PL in \textit{step 3} can be derived using the existing models defined in \textit{Table 7.4.1-1: Pathloss models} in \cite{3gpp38901}. For example, the calculation formula for the NLoS condition in the Indoor Hotspot (InH) scenario is represented as
\begin{equation}
\label{eqn_vrpl}
P L^q[\text{dB}]=-17.3-24.9 \log _{10}\left(f_c\right)-38.3 \log _{10}\left(d_{rp}^{3 \mathrm{D},q}\right)
\end{equation}
in the dB domain, where $f_c$ is the center frequency normalized by 1 GHz. The distance-related values are normalized by 1 m.

\textit{Steps 4-8} follow the communication standardization \cite{3gpp38901} to generate the other large-scale parameters, e.g., shadow fading ($SF^q$), DS, AS, and the small-scale parameters, e.g., the relative delays ($\tau_{n,m}$), AoDs ($\phi_{n,m}^{\rm{aod}}$), ZoDs ($\theta_{n,m}^{\rm{zod}}$), and Cross-Polarization Ratios (XPRs) of the multipaths in Tx$\&$Rx-RP $q$ sub-channel. These parameters form the modeling results for the ISAC monostatic background channel. Notably, since the positions of the multi-RPs and the actual Rx differ, the Azimuth angles of Arrival (AoAs) and Zenith angles of Arrival (ZoAs) of the channel multipaths generated by the standardization cannot be directly reused from the RP-generated results. Considering the single-hop characteristic of monostatic channel multipaths, a straightforward approach is to set the arrival angles equal to the departure angles, i.e., 
\begin{equation}
\label{eqn_vrang}
\phi_{n,m}^{\rm{aoa}}=\phi_{n,m}^{\rm{aod}},\ \ 
\theta_{n,m}^{\rm{zoa}}=\theta_{n,m}^{\rm{zod}}.
\end{equation}
Additionally, the arrival angles can also be flexibly configured to support modeling of potential multi-hop propagation.

\begin{figure*}[b]
\label{eqn_2}
\begin{align}
\hline \notag\\
\label{eqn_hnlos}
H_{u,s,n,m}^{{\rm{nlos}}}(\tau,t)=& { {\sqrt{\frac{P_n}{M}} {{{\left[ {
\begin{array}{*{20}{c}}
{{F_{rp,u}^\theta}({\theta _{{n},{m}}^{\rm{zoa}}},{\phi _{{n},{m}}^{\rm{aoa}}})}\\
{{F_{rp,u}^\phi}({\theta _{{n},{m}}^{\rm{zoa}}},{\phi _{{n},{m}}^{\rm{aoa}}})}
\end{array}} \right]}^T}\left[ 
{\begin{array}{*{20}{c}}
{\exp (j\Phi _{{n},{m}}^{\theta \theta })}&{\sqrt {\kappa _{{n},{m}}^{ - 1}} \exp (j\Phi _{{n},{m}}^{\theta \phi })}\\
{\sqrt {\kappa _{{n},{m}}^{ - 1}} \exp (j\Phi _{{n},{m}}^{\phi \theta })}&{\exp (j\Phi _{{n},{m}}^{\phi \phi })}
\end{array}} \right]} } }\notag\\
&\cdot 
\left[ {\begin{array}{*{20}{c}}
{{F_{tx,s}^\theta}({\theta _{{n},{m}}^{\rm{zod}}},{\phi _{{n},{m}}^{\rm{aod}}})}\\
{{F_{tx,s}^\phi}({\theta _{{n},{m}}^{\rm{zod}}},{\phi _{{n},{m}}^{\rm{aod}}})}
\end{array}} \right]\exp \left( {\frac{{j2\pi (\widehat r_{rp,{n},{m}}^{T} \cdot {{\overline d }_{u}})}}{{{\lambda _0}}}} \right)\exp \left( {\frac{{j2\pi (\widehat r_{tx,{n},{m}}^{T} \cdot {{\overline d }_{s}})}}{{{\lambda _0}}}} \right)\notag \\
&\cdot 
\exp \left( {j2\pi {f_{d,{n},{m}}}t} \right)\delta ({\tau } - {\tau _{{n},{m}}}-{d^{{\rm{3D}},q}_{rp}}/{c}-\Delta \tau).
\end{align}
\end{figure*}

Based on (\ref{eqn_hnlos}), all resolvable paths in Tx$\&$Rx-RP $q$ sub-channel at the NLoS condition at time $t$ is expressed as
\begin{equation}
\label{eqn_vrh}
H_{u, s}^q(\tau,t)=\sum\limits_{n}^{N}{\sum\limits_{m}^{M}H_{u, s,n,m}^{\mathrm{\rm{nlos}}}(\tau, t)}.
\end{equation}
In this formula, the clusters are indexed as $n=1,2,...,N$, and paths within each cluster as $m=1,2,...,M$. Moreover,
\begin{itemize}
\item{$(\cdot)^T$ stands for matrix transposition.}
\item{$\lambda_0$ is the wavelength of the carrier frequency.}
\item{$P_n$ is the power for cluster $n$.}
\item{$\theta _{{n},{m}}^{\rm{zoa}}$, $\phi _{{n},{m}}^{\rm{aoa}}$, $\theta _{{n},{m}}^{\rm{zod}}$, and $\phi _{{n},{m}}^{\rm{aod}}$ represent the ZoA, AoA, ZoD, and AoD for path $m$ in cluster $n$ at the RP and monostatic Tx, respectively.}
\item{$F_{rp,u}^\theta$, $F_{rp,u}^\phi$, $F_{tx,s}^\theta$, and $F_{tx,s}^\phi$ are the radiation patterns of the RP antenna $u$ or Tx antenna $s$, in the $\theta$ or $\phi$ polarization, respectively.}
\item{$\Phi_{{n},{m}}^{i j }$, $i,j\in \{\theta, \phi \}$, denotes the random initial phases for path $m$ in cluster $n$ for four different polarization combinations ($\theta\theta$, $\theta\phi$, $\phi\theta$, $\phi\phi$).}
\item{$\kappa _{{n},{m}}$ represents the XPRs for path $m$ in cluster $n$.}
\item{$\widehat r_{rp,{n},{m}}$ and $\widehat r_{tx,{n},{m}}$ are the spherical unit vectors of path $m$ in cluster $n$ at the RP and Tx sides, respectively.}
\item{${\overline d }_{u}$ and ${\overline d }_{s}$ are the location vectors of antennas $u$ and $s$ at the RP and Tx sides, respectively.}
\item{$f_{d,{n},{m}}$ denotes the Doppler shift for path $m$ in cluster $n$.}
\item{$\tau _{{n},{m}}$ is the delay for path $m$ in cluster $n$.}
\item{$\Delta \tau$ is the excess delay, which follows a lognormal distribution at the NLoS condition, as specified in \textit{Chapter 7.6.9} of \cite{3gpp38901}.}
\end{itemize}

In \textit{step 9}, the initial phases of multipaths are randomly generated. In \textit{steps 10} and \textit{11}, the total monostatic background channel between monostatic Tx and Rx is generated. The statistical PL and shadow fading with sub-channels comprising multi-RPs can be defined as
\begin{gather}
\label{eqn_mrppl}
PL[{\rm{dB}}]=10\cdot\log_{10}{\left(\sum_{q=1}^Q PL^q\right)},\\
SF[{\rm{dB}}]=10\cdot\log_{10}{\left(\sum_{q=1}^Q SF^q\right)}.
\end{gather}
Finally, the ISAC monostatic background CIR can be generated by the proposed channel model as the superposition of sub-channels, which is modeled as
\begin{equation}
\label{eqn_mrph}
G_{u, s}(\tau, t)=\sum_{q=1}^Q \sqrt{PL^q \cdot SF^q}\cdot H_{u, s}^{q}(\tau, t).
\end{equation}

\section{Model Parameterization and Validation}\label{section4}

\begin{figure}[h]
\centering
\includegraphics[width=2.8in]{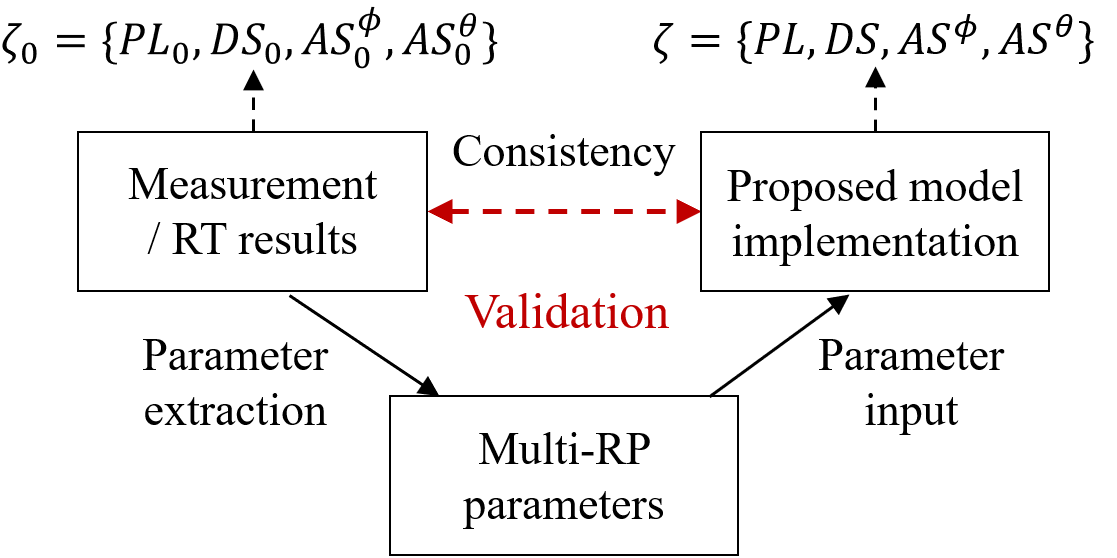}
\caption{Schematic illustration of model parameterization and validation. }
\label{fig_validation}
\end{figure}

Based on the proposed model, the complex reconstruction of the monostatic background channel is simplified to the determination of multi-RPs. In this paper, we validate the proposed model and explore the multi-RP determination method using measured indoor scenario as a benchmark. The proposed strategy illustrates in Fig. \ref{fig_validation}.

Specifically, using the obtained multi-RPs as input, the proposed channel can be implemented following the modeling procedure outlined in Section \ref{section3}-B. Then, by evaluating the consistency between the channel propagation parameters obtained from monostatic background channel measurements (${\boldsymbol{\zeta}_0}$) or RT results and those derived from the proposed model (${\boldsymbol{\zeta}}$), the multi-RP extraction is refined, while also validating the model accuracy.

In this paper, a GA-MRPE method is proposed in Section \ref{section4}-A to extract the optimal multi-RPs by minimizing the discrepancy between ${\boldsymbol{\zeta}}_0$ and ${\boldsymbol{\zeta}}$. Section \ref{section4}-B validates the iterative performance of this algorithm and presents the optimized multi-RP parameters based on the indoor measurements (as described in Section \ref{section2}). Applying the optimal and simplified model parameters, ${\boldsymbol{\zeta}}^*$ is calculated in Section \ref{section4}-C to validate the accuracy of the proposed model.

\begin{algorithm}[t]
\caption{GA based Multi-RP Extraction (GA-MRPE) Algorithm}\label{alg:alg1}
\renewcommand{\algorithmicrequire}{\textbf{Input:}}
\renewcommand{\algorithmicensure}{\textbf{Output:}}
\begin{algorithmic}[1]
\REQUIRE Maximum iterations $E$, convergence threshold $\epsilon$, population size $L$, crossover rate $R_{cro}$, mutation rate $R_{mut}$,  weights $w_1$, $w_2$, $w_3$, and $w_4$, and constraints $Q_{\rm{min}}$, $Q_{\rm{max}}$, ${d}_{\rm{min}}^{\rm{3D}}$, ${d}_{\rm{max}}^{\rm{3D}}$, ${\phi}_{\rm{min}}^{\rm{aod}}$, ${\phi}_{\rm{max}}^{\rm{aod}}$, ${\theta}_{\rm{min}}^{\rm{aod}}$, ${\theta}_{\rm{max}}^{\rm{aod}}$, $\Delta_d$, $\Delta_\phi$, and $\Delta_\theta$. Measured results $\boldsymbol {\zeta}_0$. Channel configuration parameters.
\ENSURE Optimal distances $\boldsymbol{d}_{mrp}^{{\rm{3D}}(*)}$ and angles $\boldsymbol{\phi}_{mrp}^{\rm{zod}(*)}$, $\boldsymbol{\theta}_{mrp}^{\rm{zod}(*)}$ with $1\!\times \!Q^*$ dimensions. Optimal fitness $\eta^{(*)}$. 

{\%\%\textit{{Module 1 (1-3):}} Initialize population}
\STATE{Generate initial population $\left[\boldsymbol{d}_{{mrp}}^{{\rm{3D}} ,(0)}, {\boldsymbol{\phi}}_{{mrp}}^{{\rm{aod}},(0)}, \boldsymbol{\theta}_{mrp}^{{\rm{zod}},(0)}\right]$ with $L\!\times \!Q_{\rm{max}}$ dimensions randomly, satisfying (\ref{eqn_opt}).}
\STATE{Randomly set certain values to \textit{NaN}, satisfying (\ref{eqn_opt}).\\}
\STATE{Evaluate fitness $\boldsymbol{\eta}^{(0)}$ for each individual via (\ref{eqn_opt}a).\\}

{\%\%\textit{Module 2 (4-12):} Iterative optimization\\}
\STATE{Set $e=0$. For population:\\}
\REPEAT
    \STATE{Select parents using roulette wheel selection.}
    \STATE{Perform crossover under $R_{cro}$, satisfying (\ref{eqn_opt}b)-(\ref{eqn_opt}i).}
    \STATE{Perform mutation under $R_{mut}$, satisfying (\ref{eqn_opt}b)-(\ref{eqn_opt}i).}
    \STATE{Evaluate $\boldsymbol{\eta}^{(e)}$ via (\ref{eqn_opt}a), set $\eta^{(e,*)}\!=\!\min{\boldsymbol{\eta}^{(e)}}$.\\}
    \STATE{Update population $\left[\boldsymbol{d}_{{mrp}}^{{\rm{3D}} ,(e+1)},{\boldsymbol{\phi}}_{{mrp}}^{{\rm{aod}},(e+1)},\boldsymbol{\theta}_{mrp}^{{\rm{zod}},(e+1)}\right]=
    \left[\boldsymbol{d}_{{mrp}}^{{\rm{3D}} ,(e)}, {\boldsymbol{\phi}}_{{mrp}}^{{\rm{aod}},(e)}, \boldsymbol{\theta}_{mrp}^{{\rm{zod}},(e)}\right]$.}
    \STATE{$e\leftarrow e+1$.}
\UNTIL {$\eta^{(e,*)}-\eta^{(e-1,*)}\leq \epsilon$ \textbf{or} $e>E$\\}

{\%\%\textit{Module 3 (13-14):} Obtain model parameters}
\STATE{Return $\eta^{(*)}=\eta^{(e,*)}$.\\}
\STATE{Return $Q^*$ and $\boldsymbol{d}_{mrp}^{{\rm{3D}}(*)}$, $\boldsymbol{\phi}_{mrp}^{\rm{aod}(*)}$, $\boldsymbol{\theta}_{mrp}^{\rm{zod}(*)}$ at $\eta^{(*)}$.\\}
\end{algorithmic}
\end{algorithm}

\subsection{Genetic Algorithm based Multi-RP Extraction}

To obtain the optimal multi-RPs for the parameterization of the proposed channel model, we perform mathematical representation of the problem.
Firstly, the PL calculation function  (\ref{eqn_mrppl}) is defined as $PL[\text{dB}]=\mathcal{PL}(\boldsymbol{d}_{mrp}^{\rm{3D}})$.
All multipaths are indexed as $m'=1,2,...,M* N* Q$.
The total AS and DS can be calculated using (\ref{eqn_spru}) and (\ref{eqn_sprr}), where the variable $m$ is replaced with $m'$. The power weight of path $m'$ belonging to cluster $n$ for Tx$\&$Rx-RP $q$ link is defined as
\begin{equation}
\label{eqn_mrppm}
p_{m'}=\frac{p_n}{M}\cdot \frac{PL^q}{PL}\cdot \frac{SF^q}{SF}.
\end{equation}
where ${PL^q}/{PL}$ and ${SF^q}/{SF}$ represent the proportion of $q^{th}$ sub-channel loss in the total large-scale fading in the liner domain. (Optionally, the proportion related to shadow fading can be omitted.) Due to the statistical properties of the model, DS and AS calculations cannot be easily expressed in a closed-form function. This paper defines them as $\mathcal{DS}(\boldsymbol{d}_{mrp}^{\rm{3D}})$, $\mathcal{AS}^\phi(\boldsymbol{d}_{mrp}^{\rm{3D}},\boldsymbol{\phi}_{mrp}^{\rm{aod}})$, and $\mathcal{AS}^\theta(\boldsymbol{d}_{mrp}^{\rm{3D}},\boldsymbol{\theta}_{mrp}^{\rm{zod}})$, respectively. (Note that, due to the round-trip nature of monostatic multipaths, as presented in (\ref{eqn_vrang}), only the departure angles of the multipaths are considered here.) Even for a given channel configuration, these parameters can only statistically converge to certain values, aligning with the law of large numbers. 

Then, we propose the following optimization problem: determining $Q$ multi-RPs such that the generated statistical parameters $\zeta$ matches the measured ones ($\zeta_0$) as closely as possible. This problem can be formulated as
\begin{subequations}
\label{eqn_opt}
\begin{align}
\min \limits_{\boldsymbol{d}_{mrp}^{\rm{3D}},\boldsymbol{\phi}_{mrp}^{\rm{aod}},\boldsymbol{\theta}_{mrp}^{\rm{zod}}} 
& w_1 \cdot \left(\mathcal{PL}(\boldsymbol{d}_{mrp}^{\rm{3D}})-PL_0\right)\\
&+w_2 \cdot \left(\mathcal{DS}(\boldsymbol{d}_{mrp}^{\rm{3D}})-DS_0\right) \notag\\
&+w_3 \cdot \left( \mathcal{AS}^\phi(\boldsymbol{d}_{mrp}^{\rm{3D}},\boldsymbol{\phi}_{mrp}^{\rm{aod}})-AS_0^\phi\right)\notag \\
&+w_4 \cdot \left( \mathcal{AS}^\theta(\boldsymbol{d}_{mrp}^{\rm{3D}},\boldsymbol{\theta}_{mrp}^{\rm{zod}})-AS_0^\theta\right),\notag \\
\textbf{\rm{s.t.}}\quad 
& Q_{\rm{min}} \leq Q \leq Q_{\rm{max}},\\
& d_{\rm{min}}^{\rm{3D}} \leq d_{\rm{los}}^{{\rm{3D}},i} \leq d_{\rm{max}}^{\rm{3D}}, \\
& \left|d_{\rm{los}}^{{\rm{3D}},i}-d_{\rm{los}}^{{\rm{3D}},j}\right| \geq \Delta_d, \\
& \phi_{\rm{min}}^{{\rm{aod}}} \leq \phi_{\rm{los}}^{{\rm{aod}},i} \leq \phi_{\rm{max}}^{{\rm{aod}}},\\
& \left|\phi_{\rm{los}}^{{\rm{aod}},i}-\phi_{\rm{los}}^{{\rm{aod}},j}\right| \geq \Delta_\phi, \\
& \theta_{\rm{min}}^{{\rm{zod}}} \leq \theta_{\rm{los}}^{{\rm{zod},i}} \leq \theta_{\rm{max}}^{{\rm{zod}}},\\
& \left|\theta_{\rm{los}}^{{\rm{zod},i}}-\theta_{\rm{los}}^{{\rm{zod},j}}\right| \geq \Delta_\theta, \\
&\forall i, j \in \{1, \dots, Q\}, i \neq j 
\end{align}
\end{subequations}
where $w_1$, $w_2$, $w_3$, and $w_4$ denote the weights for minimizing PL-related, azimuth AS-related, zenith AS-related, and DS-related parameters, respectively. $\Delta_d$, $\Delta_\phi$, and $\Delta_\theta$ represent the minimum spacing constraints for distance, AoD, and ZoD among multi-RPs, respectively.     

Genetic Algorithm (GA), as a classic technique, is highly suitable for global optimization \cite{mirjalili2019genetic}, making it an effective alternative for addressing the above issue. This paper proposes a GA based Multi-RP Extraction (GA-MRPE) method, the detailed implementation is demonstrated in Algorithm \ref{alg:alg1}.
In this algorithm, the maximum number of iterations for jointly optimizing the multi-RP distances and angles is defined as $E$. The remaining input parameters are then configured. In \textit{Module 1} (lines 1-3) of Algorithm \ref{alg:alg1}, the initial population with dimensions $L\times Q_{\text{max}}$ are generated. Each individual, of size $l\times Q_{\text{max}}$, contains randomly determined $\boldsymbol{d}_{{mrp}}^{{\rm{3D}} ,(0)}, {\boldsymbol{\phi}}_{{mrp}}^{{\rm{aod}},(0)}, \boldsymbol{\theta}_{mrp}^{{\rm{zod}},(0)}$, representing the positions of multi-RPs.
These parameters will be repeatedly generated until they satisfy the constraints (\ref{eqn_opt}b)-(\ref{eqn_opt}i). For each individual, a random number of elements ranging from 0 to $Q_{\rm{max}}-1$ are set to invalid (\textit{NaN}) to allow flexible determination of the multi-RP number. 
The fitness vector $\boldsymbol{\eta}^{(0)}$ with dimensions $L\times 1$ is calculated for each individual via (\ref{eqn_opt}a).

\textit{Module 2} (lines 4-12) of Algorithm \ref{alg:alg1} demonstrates the core GA-based optimization process. Specifically, the iteration number $e=0$ is initially set. For population in the $e^{th}$ iteration, parent individuals are selected through roulette wheel selection \cite{lipowski2012roulette}, where the probability of selecting the $l^{th}$ individual can be calculated as $(\max{\boldsymbol{\eta}^{(e)}}-{\eta}_l^{(e)})/\sum_l(\max{\boldsymbol{\eta}^{(e)}}-{\eta}_l^{(e)})$. The selected population undergoes single crossover, where adjacent pairs are chosen based on $R_{cro}$ and a random crossover point is selected to swap values. The offspring are then subject to mutation with $R_{mut}$. These operations are performed while ensuring that all constraints are satisfied. The fitness vector $\boldsymbol{\eta}^{(e)}$ of the mutated population is evaluated, and the minimum fitness value is recorded as $\eta^{(e,*)}$. The population is updated for the $(e+1)^{th}$ iteration. The above steps are repeated until convergence, i.e., $\eta^{(e,*)}-\eta^{(e-1,*)}\leq \epsilon$, or the iteration count $e>E$ is reached. 

In \textit{Module 3} (lines 13-14), the optimal fitness is updated to $\eta^{(*)}$, and the corresponding optimal $\boldsymbol{d}_{mrp}^{{\rm{3D}}(*)}$, $\boldsymbol{\phi}_{mrp}^{\rm{aod}(*)}$, and $\boldsymbol{\theta}_{mrp}^{\rm{zod}(*)}$ at $\eta^{(*)}$ with dimensions $1\!\times \!Q^{*}$ (At this point, the number of effective multi-RPs is $Q^{*}$.) are returned. This strategy ensures joint optimization of all statistical propagation parameters, providing a robust and accurate solution for the multi-RP extraction problem.

\begin{figure}[h]
\centering
\includegraphics[width=3.0in]{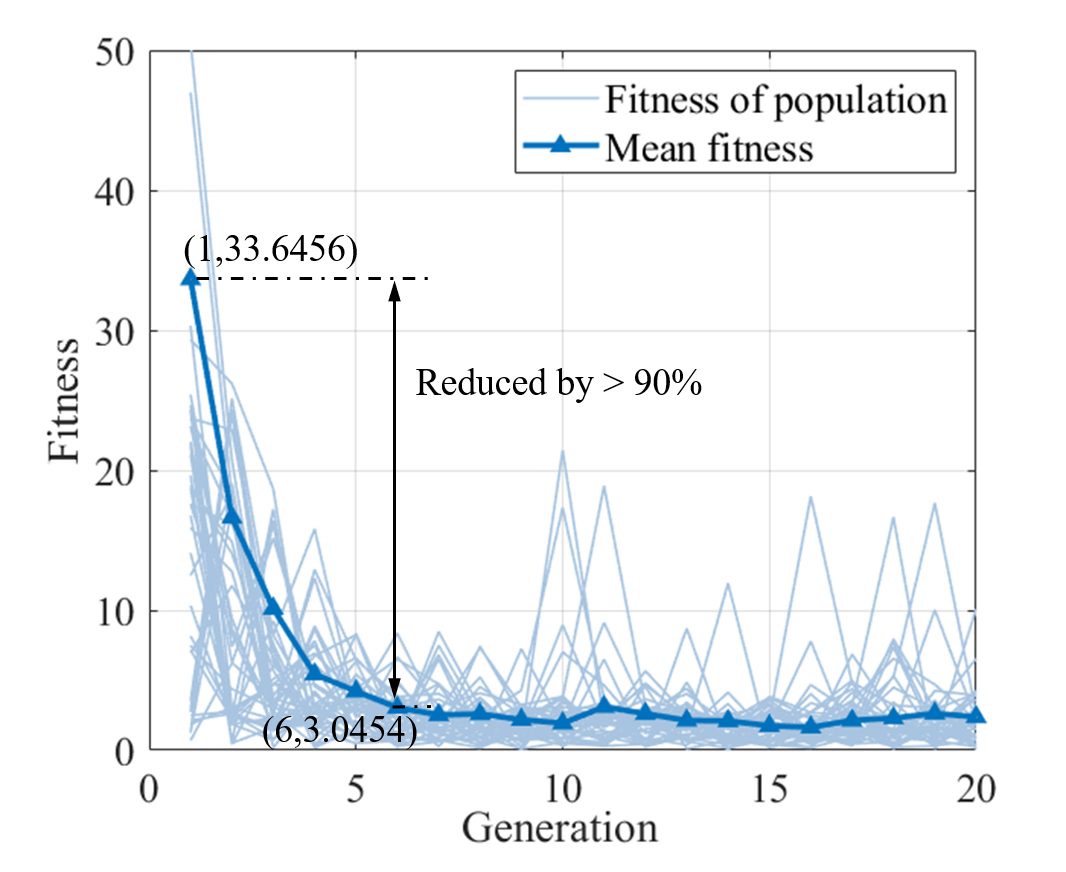}
\caption{Statistical parameter errors in the optimization process, calculated via (\ref{eqn_opt}a) and based on measurements. The light blue lines represent the error curves of 40 sets of individuals in the population over 20 iterations, while the dark blue line with triangle markers indicates the average error across the individuals (i.e., the values denoted by the light blue lines).}
\label{fig_Opt_iter}
\end{figure}

\subsection{Model Parameterization with Optimal Multi-RPs}

Next, we perform simulations based on the measurements implemented in this paper. Since the ZoD remains constant at 90$^\circ$, we optimize only the distances and AoDs of multi-RPs here, setting $w_3=0$ in (\ref{eqn_opt}a) and ignoring (\ref{eqn_opt}g) and (\ref{eqn_opt}h). In Algorithm \ref{alg:alg1}, we set the iteration-related parameters as $E=20$ and $\epsilon=10^{-6}$. The population parameters for the GA are $L=40$, $R_{cro}=100\%$, $R_{mut}=20\%$. As for the parameter weights, considering the critical role of PL in link budget evaluation, we ensure that $\mathcal{PL}(\boldsymbol{d}_{mrp}^{\rm{3D}})=PL_0$, i.e., $w_1=\infty$ here. To ensure fairness in the weighting of the remaining parameters in (\ref{eqn_opt}a), we configure $w_2=1$ and $w_4=10^{8}$, so that the error units of DS and AS correspond to 10 ns and 1$^\circ$, respectively. 

\begin{table*}[htbp]
\caption{Comparison of Measured and Modeled DS and AS Values under Different Multi-RP Configurations}\label{table_2}
    \centering
    \renewcommand{\arraystretch}{1.2}
    \begin{tabular}{cccccccc}
        \hline
        \multirow{2}{*}{Parameters} & \multirow{2}{*}{Measured results} & \multicolumn{6}{c}{Modeled results (under different RP configurations)} \\
        \cline{3-8}
        & & Optimal & 1 & 2 average & 3 average & 4 average & 5 average  \\
        \hline
        ${DS\ [ns]}$ &  32.92 & 32.96 & 24.85 & 32.13  & 33.60  & 33.61 & 35.91      \\
        $\varepsilon_{\text{norm}}$ of $DS$ & / & 0.12\% & 24.51\% & 2.40\% & 2.07\% & 2.10\% & 9.08\%\\
        ${AS^\phi\ [^\circ]}$ & 89.98 & 89.78  & 42.00 &  86.80 & 91.05  & 92.94 & 93.22 \\
        $\varepsilon_{\text{norm}}$ of $AS^\phi$ & / & 0.22\% & 53.32\% & 3.53\% & 1.19\% & 3.29\% & 3.60\%\\
        \hline
    \end{tabular}
\end{table*}

The constraint-related parameters are set as $Q_{\rm{min}}=1$, $Q_{\rm{max}}=5$, ${d}_{\rm{min}}^{\rm{3D}}=0$ m, ${d}_{\rm{max}}^{\rm{3D}}=100$ m, ${\phi}_{\rm{min}}^{\rm{aod}}=0^\circ$, ${\phi}_{\rm{max}}^{\rm{aod}}=360^\circ$, $\Delta_d=0$ m. Given the model complexity, we aim to generate the minimum number of multi-RPs while maintaining model performance. To prevent RP excessive overlap, we apply a separation is applied through 
$\Delta_\phi=20^\circ$ in the angular domain.
The first AoD angle, ${\phi}_{\rm{los}}^{\rm{aod},1}$, is fixed at 0$^\circ$, as its absolute position does not affect the AS result due to the circular and periodic nature of the angular domain.
The measurement results $\boldsymbol {\zeta}_0=\{-80.8125,3.292\times 10^{-8},89.98,NaN\}$ [dB,s,$^\circ$,NaN] are calculated in Section \ref{section2}-B. Additionally, as for the channel parameters, the Indoor Hotspot (InH) scenario is adapted, with a center frequency of 28 GHz and a system bandwidth of 1 GHz, consistent with our measurement campaign.

\begin{figure}[h]
\centering
\subfloat[]{\includegraphics[width=3in]{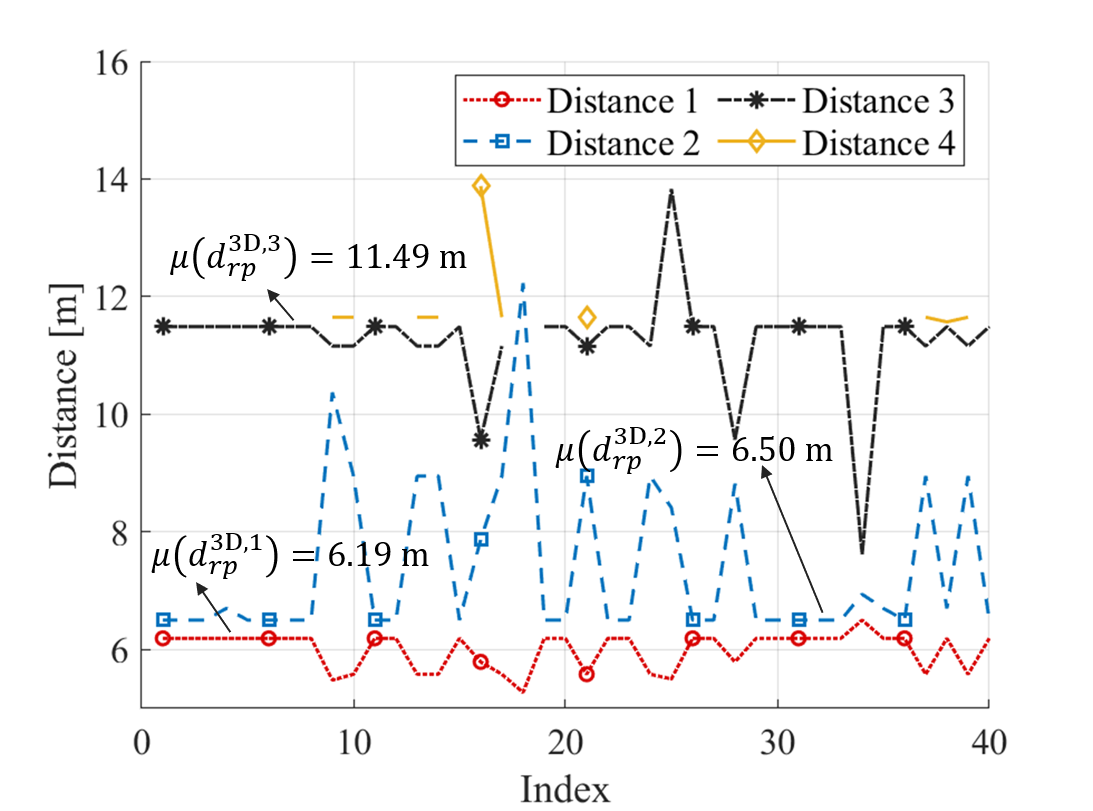}}\\
\vspace{-0.3cm}
\subfloat[]{\includegraphics[width=3in]
{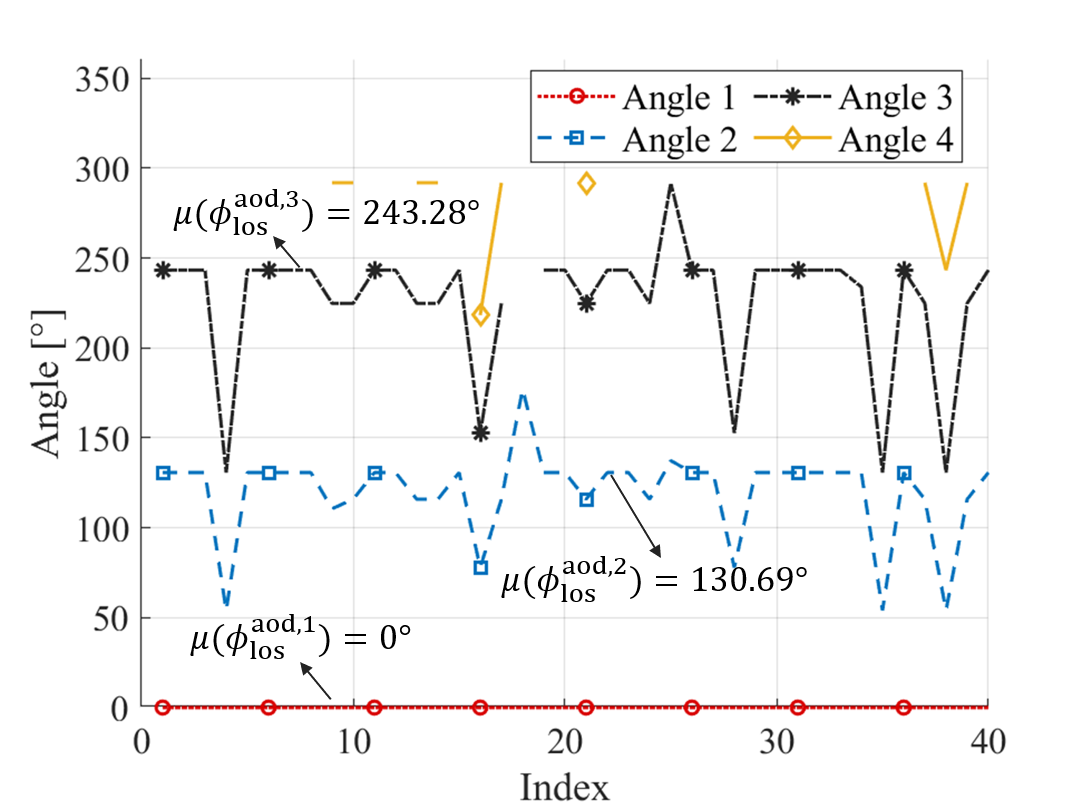}}\\
\vspace{-0.1cm}
\caption{Optimization results showing (a) 3D distances and (b) AoDs for the top 5\% individuals (40 in total) with the lowest parameter errors.}
\label{fig_Opt_results}
\end{figure}

Fig. \ref{fig_Opt_iter} shows the variations in statistical parameter error (i.e., the fitness of GA-MRPE) over 20 iterations. As the number of iterations increases, the parameter error gradually decreases, and GA-MRPE algorithm demonstrates good convergence. The optimal multi-RP parameters are determined to be $Q^*=3$, $\boldsymbol{d}_{mrp}^{{\rm{3D}}(*)}=\{6.19,6.50,11.49\}$ m, $\boldsymbol{\phi}_{mrp}^{\rm{aod}(*)}=\{0,130.69,243.28\} ^\circ$. To further assess the optimization process, the top 5\% of individuals with the lowest errors are selected, and their AoDs and 3D distances are presented in Fig. \ref{fig_Opt_results}(a) and (b), respectively. It can be observed that most of the results correspond to $Q^*=3$, and the values fluctuate around the optimal solution. 

\begin{figure*}[!h]
\centering
\subfloat[]{\includegraphics[width=2.5in]{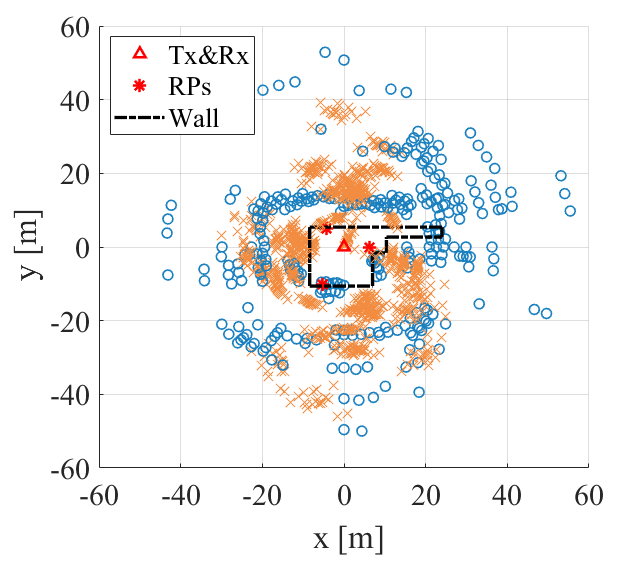}}
\hspace{2mm}
\subfloat[]
{\includegraphics[width=2.6in]{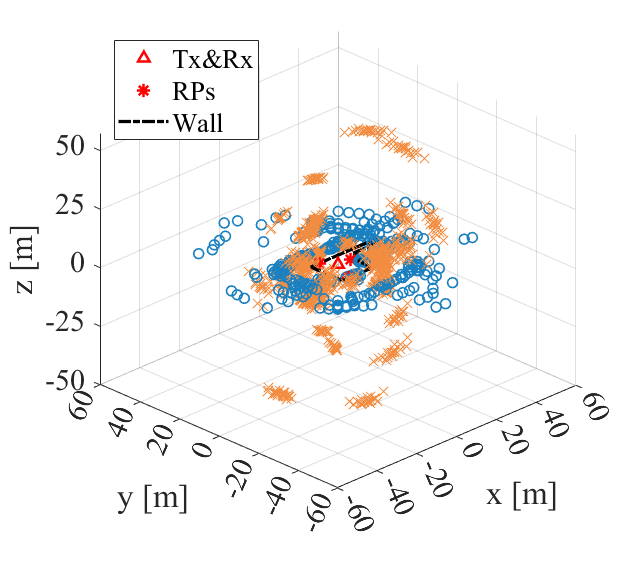}}\\
\subfloat[]{\includegraphics[width=2.6in]{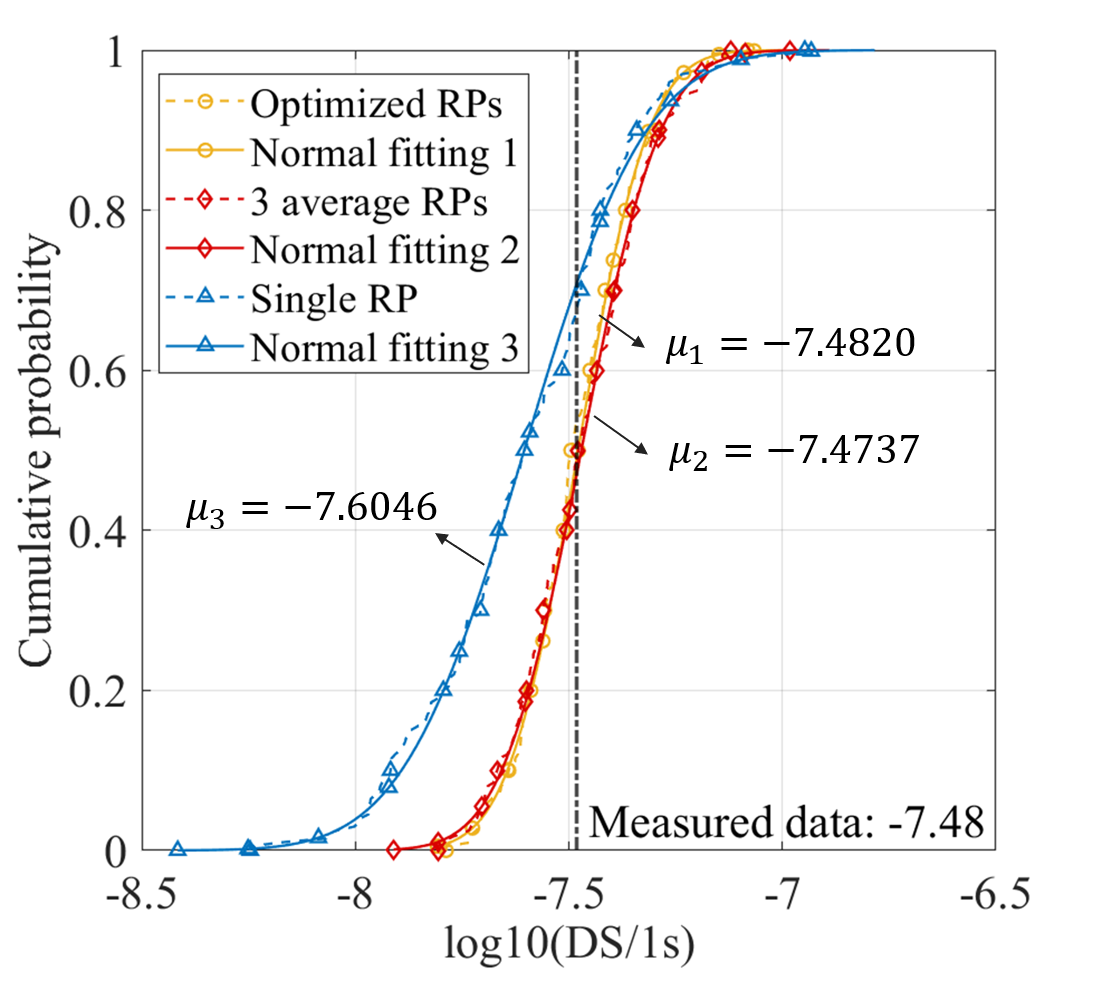}}
\subfloat[]{\includegraphics[width=2.6in]{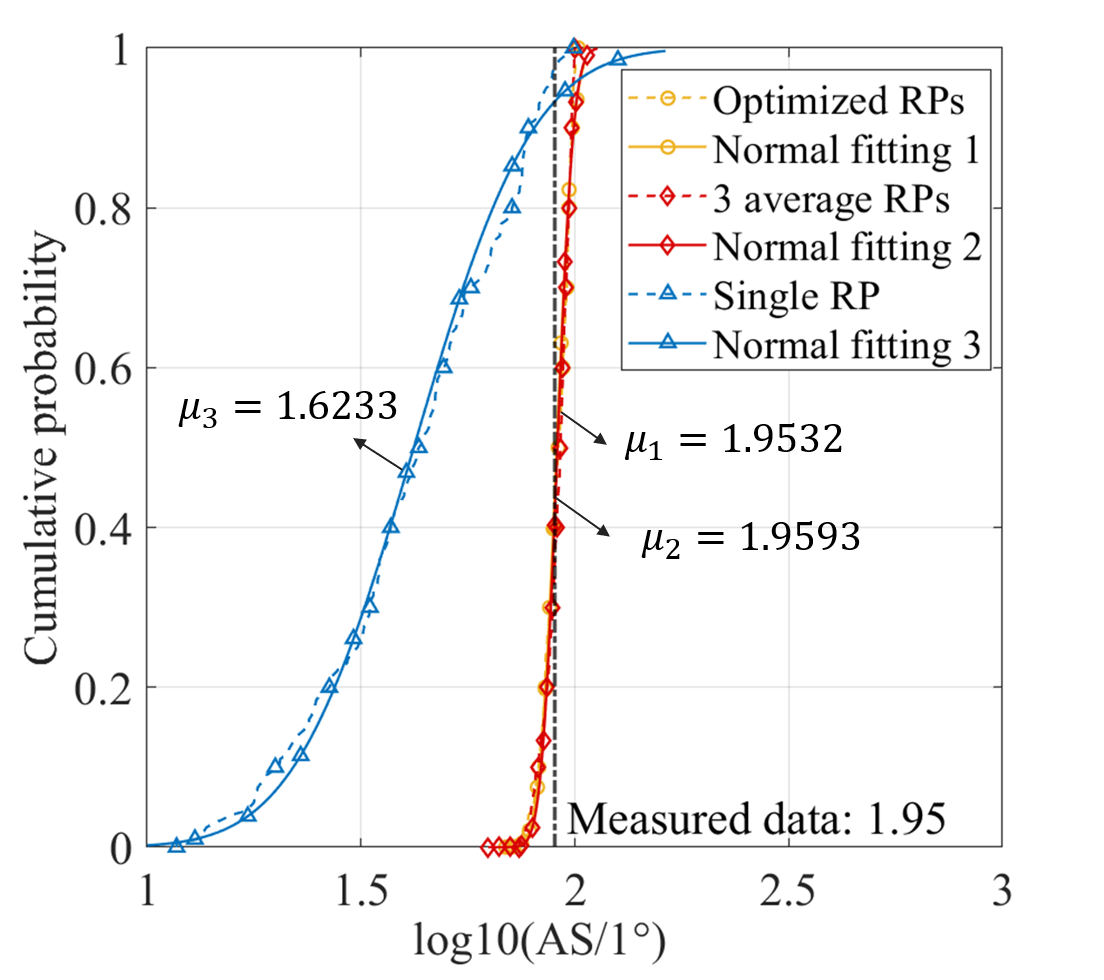}}\\
\caption{The (a) 2D and (b) 3D layouts of the measured and modeled paths under single simulation. The blue dots represent the effective paths from measured monostatic background channel, and the orange crosses indicate the generated paths based on the proposed multi-RP channel model. Simulated CDFs of modeled channel parameter under 200 simulations, where (c) $\rm{log}_{10}(DS/1s)$ and (d) $\rm{log}_{10}(AS/1^\circ)$ of AoDs with optimized multi-RPs, simplified 3 average multi-RPs, and single RP (i.e., communication configure). The black lines with -7.48 and 1.95 mark the corresponding measured values 32.92 ns and 89.98$^\circ$, which serves as a baseline for evaluating the accuracy of the proposed model.}
\label{fig_simmrp}
\end{figure*}

\subsection{Model Validation Using Measured Parameters}

Based on the obtained optimal multi-RP parameters, the ISAC monostatic background CIR is generated following the procedure outlined in Section \ref{section3}. To validate the model’s performance, Fig. \ref{fig_simmrp}(a) and (b) show the 2D and 3D layouts of the measured and modeled paths in a single channel simulation. The path positions are determined based on the departure angles and propagation delays. In these figures, the monostatic Tx$\&$Rx and multi-RPs are represented by red triangles and stars, respectively. The black dashed boxes illustrate the wall layout of the measurement environment, corresponding to Fig. \ref{fig_sce}(a). The blue dots represent the effective paths from the measured monostatic background channel, corresponding to Fig. \ref{fig_padp}. Due to the round-trip nature of the monostatic channel paths, the contour indicated by the blue dots is approximately twice the distance of the black box. The orange crosses represent the generated paths by the proposed multi-RP channel modeling. It can be observed that the modeled multipaths (orange crosses) effectively cover the discrete region corresponding to the measured multipaths (blue dots). 

Due to the stochastic foundation of the GBSM, the extracted multi-RP parameters in the proposed statistical channel model are expected to exhibit statistical variability, as shown in Fig. \ref{fig_Opt_results}. With maintaining satisfactory modeling performance, the proposed model can be appropriately simplified based on the optimal results to achieve lower complexity. For example, with $Q^*=3$, Fig. \ref{fig_Opt_results}(a) shows that the first two distances are nearly identical, while the third is nearly twice as large and contributes less power, thus having limited impact on the DS. This suggests that when the multi-RPs are nearly equal in distance, a good DS can still be achieved, and in this case, the number of RPs has minimal impact on the DS. Therefore, a simplified approach allows setting the three distances equal while maintaining the PL constraint. In our measurements, the corresponding 3D distance of 3 average RPs is $\boldsymbol{d}_{mrp}^{{\rm{3D}}}=\{6.95,6.95,6.95\}$ m. As shown in Fig. \ref{fig_Opt_results}(b), the three optimal AoDs are approximately evenly distributed, allowing us to set $\boldsymbol{\phi}_{mrp}^{\rm{aod}}=\{0,120,240\} ^\circ$. 

Based on the above analysis of the optimal and average multi-RPs for $Q^* = 3$, we approximate the model results of average multi-RPs as the represented values for a given $Q\in\{1,2,3,4,5\}$ to assess the impact of the number of multi-RPs on the model's accuracy. When $Q=1$, the corresponding 3D distance is $\boldsymbol{d}_{mrp}^{{\rm{3D}}}=\{5.22\}$ m (satisfying the PL equivalence condition), and the AoD is $\boldsymbol{\phi}_{mrp}^{\rm{zod}}=\{0\} ^\circ$. Moreover, $\boldsymbol{d}_{mrp}^{{\rm{3D}}}=\{6.25,6.25\}$ m and $\boldsymbol{\phi}_{mrp}^{\rm{zod}}=\{0,180\} ^\circ$ for $Q=2$, $\boldsymbol{d}_{mrp}^{{\rm{3D}}}=\{7.49,7.49,7.49,7.49\}$ m and $\boldsymbol{\phi}_{mrp}^{\rm{zod}}=\{0,90,180,270\} ^\circ$ for $Q=4$, and $\boldsymbol{d}_{mrp}^{{\rm{3D}}}=\{7.94,7.94,7.94,7.94,7.94\}$ m and  $\boldsymbol{\phi}_{mrp}^{\rm{zod}}=\{0,72,144,216,288\} ^\circ$ for $Q=5$. Then, we conduct 200 model simulations based on both the optimal multi-RPs and simplified average multi-RPs. The simulated mean values of the DS and AS in the proposed model are listed in Table. \ref{table_2}. It can be observed that with an increasing number of multi-RPs, the simulated DS and AS values progressively increase, indicating a more discrete distribution of multipaths in both the angular and delay domains. To further validate the accuracy of the proposed ISAC multi-RP channel model, we calculate the normalized error ($\varepsilon_{\text{norm}}$) between the modeled and measured parameters as
\begin{equation}
\label{eqn_mrppm}
\varepsilon_{\text{norm}}=\frac{\left|v_0-{v}\right|}{v_0} \times 100\%, v\in \{DS\ \text{or}\ AS^\phi\}.
\end{equation}
When $Q=1$, i.e., when reusing the communication standardized parameters, the results show significant deviation from the measured DS and AS values, with errors of 24.51\% and 53.32\%, as shown in Table. \ref{table_2}. However, for the average RP configurations at $Q=2,3,4,5$, the modeled values are relative closer to the measured values. (Note that if the multi-RPs are not evenly distributed, the results may be worse.) The mean DS and AS values from the optimal multi-RP configuration exhibit the closest match to the measured values, with errors reduced by more than 20\% and 50\% compared to the single RP (communication configuration) results, specifically 0.12\% and 0.22\%, respectively.

The simulated CDFs of $\rm{log}_{10}(DS/1s)$ and $\rm{log}_{10}(AS/1^\circ)$ for the optimal multi-RPs and simplified 3 average multi-RPs are shown in Fig. \ref{fig_simmrp}(c) and (d), respectively. For comparison, the simulation result for $Q=1$, based on communication parameters \cite{3gpp38901, jiang2024novel}, is also included. All these simulated CDFs can be well fit by normal distributions. It can be observed that the proposed modeling framework, utilizing either the optimal or the simplified 3 average multi-RPs, effectively reproduces the stochastic propagation parameters of the measured ISAC monostatic background channel. In future work, by parameterizing and generating the multi-RPs, which follow parameterized distributions for typical ISAC scenarios, we will be able to realize any given channel. This model demonstrates satisfactory accuracy and holds potential for supporting 6G ISAC standardization.

\section{Conclusion}\label{section5}

This paper proposes a novel multi-RP modeling framework to characterize ISAC monostatic background channel, which is parametrized and validated based on realistic measurements. Firstly, an indoor channel measurement campaign at 28 GHz reveals that the effective multipaths in the ISAC monostatic background channel exhibit pronounced single-hop propagation and a discrete distribution. The extracted channel propagation parameters, including PL, DS, and azimuth AS, provide a data foundation for model development. 
Inspired by observed channel properties, each monostatic background path is mapped to a virtual anchor, with nearby ones represented by a sub-channel from a communication Rx-like RP. Then, this paper proposes a novel multi-RP based stochastic channel model, which characterizes the ISAC monostatic background channel as the superposition of sub-channels between the monostatic Tx$\&$Rx and multi-RPs. A 3GPP-extended implementation framework is also introduced, ensuring standardization compatibility and enabling low-complexity realization of ISAC monostatic background channel. To support model parameterization, a GA-MRPE algorithm is developed to extract optimal multi-RP configuration by minimizing the discrepancy between practical and modeled channel parameters. As the number of averaged multi-RPs increases, the simulated DS and AS values in the proposed model progressively increase, indicating a more discrete multipath distribution in both the angular and delay domains. Based on the measurements, the optimal number of multi-RPs is determined to be 3, with evenly distributed angles and equal distances considered as a simplification. Results validate that the proposed model accurately matches the measured ISAC channel, with DS and AS errors reduced by more than 20\% and 50\%, respectively, compared to reusing communication standardized parameters. The proposed model demonstrates satisfactory accuracy in characterizing the ISAC monostatic background channel, which facilitates sensing performance evaluation and contributes to the advancement of 6G ISAC standardization.

\appendices
\section{List of Abbreviations}

Table \ref{table_abb} lists the acronyms used in this paper, arranged in alphabetical order.

\begin{table}[h]
\caption{List of Abbreviations}\label{table_abb}
\centering
\begin{tabular}{cc}
\hline
Abbreviation & Values\\
\hline
2D / 3D & Two / Three-Dimensions\\
3GPP & Third Generation Partnership Project\\
4G / 5G / 6G & Fourth / Fifth / Sixth Generation\\
AS & Angular Spread\\
AoA & Azimuth angle of Arrival\\
AoD & Azimuth angles of Departure\\
BPSK & Binary Phase Shift Keying\\
BS & Base Station\\
CDF & Cumulative Distribution Function\\
CIR & Channel Impulse Response\\
DS & Delay Spread\\
GA & Genetic Algorithm\\
GA-MRPE & GA-based Multi-RP Extraction\\
GBSM & Geometry-Based Stochastic channel Model\\
GCS & Global Coordinate System\\
InH & Indoor Hotspot\\
ISAC & Integrated Sensing And Communication \\
ITU & International Telecommunication Union\\
LNA & Low Noise Amplifier\\
LoS & Line-of-Sight \\
mmWave & millimeter wave\\
NLoS & Non-Line-of-Sight \\
PADP & Power-Angle-Delay Profiles\\
PL & Path Loss\\
PN & Pseudo Noise\\
RAN & Radio Access Network\\
RAN1 & RAN Working Group 1\\
RMS & Root Mean Square\\
RP & Reference Point\\
RT & Ray-tracing\\
Rx & Receiver \\
SNR & Signal-to-Noise Ratio \\
ST & Sensing Target \\
TR & Technical Report\\
Tx & Transmitter \\
UT & User Terminal\\
XPR & Cross-Polarization Ratio\\
ZoA & Zenith angle of Arrival\\
ZoD & Zenith angle of Departure\\
\hline
\end{tabular}
\end{table}

\bibliography{reference}

\end{document}